\documentclass{aa}
\usepackage{graphicx}
\usepackage{txfonts}
\usepackage{natbib}
\usepackage[
  breaklinks = true,
   colorlinks = true,
  urlcolor   = blue,
  citecolor  = blue,
  linkcolor  = blue,
]{hyperref}
\graphicspath{ {./figures/} }

\begin{document}

\title{
The role of small-scale surface motions in the transfer of twist to a solar jet from a remote stable flux rope
}

\author{Reetika Joshi
\inst{1,2}\and
Brigitte Schmieder
\inst{1,3,4}\and
Guillaume Aulanier
\inst{1}\and
V\'eronique Bommier
\inst{1}\and
\and
Ramesh Chandra
\inst{2}
}
\institute{LESIA, Observatoire de Paris, Universit\'e PSL, CNRS, Sorbonne Universit\'e, Universit\'e Paris Diderot, 5 place Jules Janssen, 92190 Meudon, France
\and
Department of Physics, DSB Campus, Kumaun University, Nainital -- 263 001, India\\
\email{reetikajoshi.ntl@gmail.com, reetika.joshi@obspm.fr}
\and
Centre for mathematical Plasma Astrophysics, Dept. of Mathematics, KU Leuven, 3001 Leuven, Belgium
 \and
School of Physics and Astronomy, University of Glasgow, Glasgow G12 8QQ, UK
}
\authorrunning{Reetika Joshi et al.} 
   \titlerunning{Transfer of twist to a jet from  a remote fluxrope}
\abstract
{
Jets often have a helical structure containing ejected plasma that is both hot and also cooler and denser than the corona. Various mechanisms have been proposed to explain how jets are triggered, primarily attributed to a magnetic reconnection between the emergence of magnetic flux and  environment or that of twisted photospheric motions that bring the system into a state of instability.
}
{
Multi-wavelength observations of a twisted jet observed with the {\it Atmospheric Imaging Assembly} (AIA) onboard the {\it Solar Dynamics Observatory} (SDO) and the {\it Interface Region Imaging Spectrograph} (IRIS) were used to understand how the twist was injected into the jet, thanks to the IRIS spectrographic slit fortuitously crossing the reconnection site at that time.
}
{
We followed the magnetic history of the active region based on the analysis of the {\it Helioseismic and Magnetic Imager} (HMI) vector magnetic field computed with the UNNOFIT code. The nature and dynamics of the jet reconnection site are characterised by the IRIS spectra.
}
{
This region is the result of the collapse of two emerging  magnetic fluxes (EMFs) overlaid by arch filament systems that have been well-observed with AIA, IRIS, and the {\it New Vacuum Solar Telescope} (NVST) in H$\alpha$.
In the magnetic field maps, we found evidence of the pattern of a long sigmoidal flux rope (FR) along the polarity inversion line between the two  EMFs, which is the  site of  the reconnection. Before the jet, an extension of the FR was present and a part of it was detached and formed a small bipole with a bald patch (BP) region, which dynamically became an X-current sheet over the dome of one EMF where the reconnection took place. At the time of the reconnection, the Mg II spectra exhibited a strong extension of the blue wing that is decreasing  over a distance of 10 Mm (from -300 km  s$
^{-1}$ to a few  km  s$^{-1}$). This is  the signature of the transfer of the twist to the jet.
}
{
A comparison with numerical magnetohydrodynamics (MHD) simulations confirms the existence of the long FR. 
We conjecture that there is a transfer of twist to the jet  during the extension of the FR to the reconnection site without FR eruption. The reconnection would start in the low atmosphere in the BP reconnection region and extend at an X-point along the current sheet formed above.}

\maketitle 
\section{Introduction} 
\label{intro}
In the 1990s, jets had already been
observed across all   temperature ranges (10$^4$ K to 10$^7$ K) in multi-wavelength observations from H$\alpha$
with ground-based instruments \citep{Gu1994,Schmieder1994b,Schmieder1995,Canfield1996} to X-rays with Yohkoh \citep{shibata1992}.
More recently, many jets have been detected in the extreme ultraviolet (EUV)  with the {\it Atmospheric
Imaging Assembly} \citep[AIA,] [] {Lemen2012} onboard  the \emph{Solar Dynamics Observatory} \citep[SDO,] [] {Pesnell2012} and with the \emph{Interface Region Imaging Spectrograph} \citep [IRIS,] [] {Pontieu2014}.
Their characteristics vary in wide parameter ranges, such as velocity (100 to 400 km s$^{-1}$),
extension (10$^{4}$ to  10$^{6}$ km), and width (10$^{4}$ to  10$^{5}$ km) \citep{Nistico2009, Schmieder2013,Joshi2017b}.

Exhibiting a twist or rotation is a common property for jets overall  \citep{Raouafi2016}. The twist of the jet may be due to  helical motions
\citep{Nistico2009,Patsourakos2008}. Twisting motions have been found  in  a large velocity range of jets or surges \citep{Chen2012,Hong2013,Zhang2014}.
In the study done by  
\citet{Schmieder2013}, a jet analysis
revealed a striped pattern of dark and bright strands propagating along the jet, as well as apparent damped oscillations across the jet. They concluded that this
is suggestive of a (un)twisting motion in the jet, possibly an Alfv\'en wave. Spectroscopic  data also provide  signatures for detecting the twist in  jets.  
For example, blue and red shifts observed along the axis of a jet in H$\alpha$ as well as in Mg II lines  were interpreted as confirmation of the existence of twist along the  jet \citep{Ruan2019}.

The   concept behind the existence of coronal  jets is based on magnetic reconnection. 
This  was  first proposed by \citet{Heyvaerts1977} and simulated by \citet{Shibata1982}, \citet{shibata1992} and \citet{yokoyama1996}. 
Magnetic reconnection has explained the  X-ray jets observed by Yohkoh \citep{Shimojo1996,Shimojo2000,Asai2001}.
It also supports 
the interpretation of jets 
observed in multi-wavelengths from X-ray \citep{Shimojo1996,Shimojo2000,Asai2001} to EUV \citep{Schmieder1983,Schmieder1988,Schmieder1994b,Chae1999,Nistico2009,Chandra2017}.
Spectroscopic and imaging  observations of small-scale events reveal bidirectional flows in transition  region lines at the jet base  which could correspond to an explosive reconnection \citep{Li2018,Ruan2019}.
There are different conditions for the  magnetic configuration of an active region (AR) to  trigger magnetic reconnection.  We may quote three types of conditions: magnetic flux emergence \citep{Archontis2004,Archontis2005,Moreno2008,Torok2009,Moreno2013},   magnetic flux cancellation \citep{Priest2018,Syntelis2019}, and magnetic instability \citep{Pariat2010,Pariat2015,Pariat2016}. The first two mechanisms predict hot and cool jets simultaneously. 
However, the presence of surges and jets is
 not frequently reported. 
 Some papers report on the X-ray jets observed by Yohkoh and associated with a surge  \citep{Schmieder1995,Canfield1996,Ruan2019}.
Radiative magnetohydrodynamics (MHD) simulations based on flux emergence  \citep{Nobrega2016,Nobrega2017} as well as the flux cancellation model 
\citep{Syntelis2019} show that surges can exist at the same time with hot jets. The cool plasma is advected  over the emergence domain without passing near the reconnection site and then flows  along the reconnected magnetic field lines. These models fit with the observations of X-ray jets { observed with} Hinode and with H$\alpha$ jets from  the Swedish 1-m Solar Telescope (SST) \citep{Nobrega2017}.  Recently, \citet{Joshi2020} presented a case-study of collimated hot jets and associated cool surges which fit in perfectly with the simulation of jets formed by flux emergence. The double-chambered structure found in the observations corresponds to the cool and hot loop regions found under the reconnection site in the models of \citet{Moreno2008}.
Jets can also be the consequence of magnetic instability. Firstly, we see there is the storage of energy  by twisted field lines or shearing of polarities. Suddenly, the system becomes unstable with a disruption of the field lines via reconnection occurs and material expelled along the open magnetic field lines. This mechanism is based on a twisted flux rope (FR) formation during the shear in the jet region  \citep{Yeates2011,Pariat2015,Pariat2016,Raouafi2016}.  The slow reconnection is driven by the response of the system to continual stressing of the closed field injecting magnetic helicity.  When the kink instability encompasses the entirety of the closed field region, it leads to a large jet eruption. In the model of \citet{Wyper2019}, the overlying magnetic field is, in fact, expelled by a gentle reconnection above the closed AR via   a breakout mechanism before the instability occurs. \citet{Pariat2015}, and \citet{Wyper2019} show the importance of the inclination of  jets favouring the jet onset for $\theta$ = 0 - 20 degrees. 
These models are based on the instability of the system; a FR formed by shear  under the reconnection point is the trigger of the helical jet. However, based on several observations, it becomes clear that the twist is not present before the reconnection but the twist of the jet is transferred during the reconnection. For example, in  \citet{Ruan2019}, the twist was transferred from twisted overlying  magnetic field lines remnant of the eruption of a filament two hours before the onset of the jet.
The null-point is the favourable location for the occurrence of magnetic reconnection.
 
 \citet{Wyper2019} recently showed that reconnection can be in a region where the magnetic  field lines are tangent to the photosphere. 
This kind of region is called
bald patch (BP) region. It  favours reconnection as a mechanism for initiating jets and surges \citep{Mandrini2002,Chandra2017, Zhao2017}. 
In these studies, the magnetic topology was derived by linear force free field (LFFF) or non linear force free field (NLFFF) magnetic field  extrapolations in the corona \citep{Mandrini2002,Chandra2017} or by directly analysing    the   observed magnetic field vector maps \citep{Bernasconi2002,Zhao2017}.  
The  occurrence of the reconnection was clearly  taking place in the  BP regions.
It was also recently  proposed that the trigger of jets can be due to the eruption of mini-filament at the jet base \citep{Sterling2016}. That model fits well with the blowout jets where the  entire region below the dome of reconnection is expelled during the eruption \citep{Moore2010}.

In this paper, 
we present   observations of a twisted jet, a surge, and a mini-flare which occurred on 22 March 2019  at 02:05 UT observed in multi-wavelengths with AIA, IRIS instruments, and with the \emph{New Vacuum Solar Telescope} \citep[NVST,] [] {NVST2014} in Fuxian Solar Observatory of Yunnan Astronomical Observatory, China  (\url{http://fso.ynao.ac.cn/datashow.aspx?id=2296}) (Sect. \ref{AIA} and \ref{IRIS}).
The IRIS spectra at the reconnection site  display 
a tilt and gradient in the spectra along  the jet base indicating the formation of  a rotating structure during the reconnection (Sect. \ref{comparison}). In Sect. \ref{reconnection}, we analyse the magnetic topology of the AR.
This region  with the mini-flare was recently
studied  
by an other group \citep{Yang2020} and independently of our study.
They performed a NLFFF extrapolation to show that the region has a fan-spine magnetic topology
with two FRs.
They proposed a break-out model which might remove the overlaying arcades, leaving a small FR to erupt and turn into a blowout jet -- as in the scenario of \citet{Sterling2016} and in the model of \citet{Wyper2017}. 

In Sect. \ref{sec_vector}, 
 our detailed consideration of the EUV data based on the analysis of the photospheric vector magnetic field obtained with the 12 min cadence \emph{Helioseismic and Magnetic Imager} \citep [HMI,] [] {Schou2012} leads us to a different interpretation for the event.
 The clues that support our interpretation are the identification of a non-eruptive FR, from which some twist is carried away, eventually reconnecting into the jet at the  `X' point of the current sheet that is dynamically formed from a BP current sheet underneath. The presence of the FR is supported by a relatively good accordance that can be seen for the patterns of transverse fields and vertical current densities, as observed by HMI, and appearing without being constrained a priori in an MHD simulation of non-eruptive FR formation with flux-cancellation of sheared loops \citep{Zuccarello2015}, carried out with the Observationally driven High-order scheme Magnetohydrodynamic{\it } (OHM) code \citep{Aulanier2005,Aulanier2010} (Sect. \ref{OHM}). The transport of twist away from the FR towards the 
 `X' point is supported by HMI observations of a moving negative flux-concentration whose transverse fields point towards a positive one.
In Sect. \ref{dis}, we discuss the two interpretations and conclude our scenario based on HMI and IRIS  observations.


\section{Instruments}
\subsection{AIA}
\label{AIA} 
In the active region (AR) NOAA 12736, a  jet as well as a surge is well-observed in the multi-wavelength  filters of AIA aboard SDO. The AIA data consists of a sample of filters with passbands  centered at different lines emitting at different temperatures from  304 \AA
(He II, T$\sim$0.05 MK), 171 \,\AA (Fe IX, T $\sim$0.6 MK), and 193 \,\AA (Fe XII, T $\sim$0.6 MK) to hotter temperatures of  94 \,\AA{} (Fe XVIII, T$\sim$6.3 MK), 131
\,\AA (T$_1$= 10 MK and T$_2$= 0.64 MK), and 211 \,\AA (T$_1$ $\sim$ 20 MK and T$_2$ $\sim$1.6 MK).
The AIA data cadence in these channels is 12 s and the
pixel size is 0.6$ ^{\prime\prime}$. 

\subsection{HMI} \label{obs_HMI}
 The longitudinal magnetic field is provided by the HMI team with a cadence of 45 s and a pixel size of 0.5$\arcsec$. To obtain the magnetic field vectors in full, we inverted the HMI level-1p IQUV data, averaged on a  12 minute cadence, by applying the Milne-Eddington inversion code UNNOFIT \citep{Bommier2007}. We selected  a large  area   covering the AR 12736 and applied a solar rotation compensation to select the same region over more than six hours of observation. We thus treated 22 maps from  21 March 2019 at 23:00 UT to  22 March 2019 at 03:12 UT and three later maps of the same region from 05:00 UT to 05:24 UT. The specificity of UNNOFIT is that a magnetic filling factor is introduced to take into account the unresolved magnetic structures as a free parameter of the Levenberg-Marquardt algorithm that fits the observed set of profiles with a theoretical one. However, for further application, we used only the averaged field, that is, the product of the field with the magnetic filling factor, as recommended by \citet{Bommier2007}. The interest of the method lies in a better determination of the field inclination. After the inversion, the 180$^{\circ}$ remaining azimuth ambiguity was resolved by applying the ME0 code developed by Metcalf, Leka, Barnes, and Crouch \citep{Leka2009} and available at \url{http://www.cora.nwra.com/AMBIG/}. After resolving the ambiguity, the magnetic field vectors were rotated into the local reference frame, where the local vertical axis is the $o_z$ axis.

\subsection{NVST}
The H$\alpha$ observations were taken   with  the NVST telescope, pointed at the AR  12736 at N09 W60 on  22 March 2019 from 00:57:00 UT to 04:37:00 UT. We used the line--center H$\alpha$ observations at 6562.8 \AA \ that were obtained in a field of view 
(FOV) of 126$\arcsec$ $\times$ 126$\arcsec$ with a cadence of 29 seconds. 
The
H$\alpha$ movie for these observations  is available at \url{http://fso.ynao.ac.cn/datashow.aspx?id=2296}. It displays  the H$\alpha$ fine structures and  shows their evolution very well.
For the current analysis, we used the level 1+ data provided by the NVST team. To focus on the jet region, we cut the data cube after rotating it  with north upwards, as in the space data (AIA  and HMI observations) and used the data of FOV of  65$\arcsec$ $\times$ 65$\arcsec$.

\subsection{IRIS} 
\label{IRIS}
The IRIS FOV was focused on AR NOAA 12736 and the pointing of
the telescope was at 709$^{\prime}$$^{\prime}$, 228$^{\prime}$$^{\prime}$
with a FOV of
60$^{\prime}$$^{\prime}$ $\times$68$^{\prime}$$^{\prime}$ 
for the slit-jaw images (SJIs).  
{The observational characteristics}
are presented in Table \ref{tab:table1}.  We used the   1330 \AA\ and 
2796 \AA\  SJIs  for this
study. 
There was no data for the   IRIS SJI  Si IV 1400 filter.
The SJI 1330 \AA\ includes the C \footnotesize{II}
\normalsize line formed at T= 30000 K, and the SJI 2796\,\AA{}
emission mainly comes from the Mg \footnotesize{II} \normalsize k
line. The Mg \footnotesize{II} \normalsize h and k lines
are formed at chromospheric temperatures, that is, between 8000 K and
15000 K \citep{Pontieu2014, Alissandrakis2018}.  The co-alignment
between the different optical channels of IRIS was achieved
by using the {\it drot\_map} in solar software to correct the differential
rotation.  The SJIs in the broadband filters (1330\,\AA{}, and
2796\,\AA{})
 were taken at a cadence of 14 s. 
 
 With IRIS, medium  coarse rasters  of four steps were performed from 01:43:27 UT to
02:42:30 UT on 22 March 2019. 
The raster
step size is 
2$^{\prime}$$^{\prime}$ so that each spectral raster spans a field of view
(FOV) of 6 $^{\prime}$$^{\prime} \times$ 62
$^{\prime}$$^{\prime}$.  The nominal spatial resolution is 0.$^{\prime}$$^{\prime}$33. 
IRIS  provides line profiles in  Mg II k 2796.4\,\AA{} and  Mg II h 2803.5\,\AA{},  Si IV (1393.76 \AA, 1402.77 \AA) and C II  (1334.54 \AA, 1335.72 \AA) lines  along four slit positions  (200 pixels) and repeated 250 times.
Calibrated level 2 data were used in this study. 
Dark current subtraction, flat field
correction, and geometrical correction have been taken into account in
the level 2 data \citep{Pontieu2014}.

\begin{table} 
\caption{
IRIS observation of AR NOAA 12736 on 22 March 2019.}
\bigskip
\centering
\setlength{\tabcolsep}{6pt}
\begin{tabular}{llll}
\hline
\hline
Location& Time & Raster& SJI\\
 & (UT) &  &
\\
\hline
x=709$\arcsec$& 01:43 to &  FOV: 6$\arcsec$ $\times$ 62$\arcsec$& FOV: 60$\arcsec$ $\times$ 68$\arcsec$\\
 y=228$\arcsec$& 02:42 &  Steps: 4 $\times$ 2$\arcsec$& C II 1330 \AA \\
 & &  Spatial & Mg II 2796 \AA \\
 & &  Resolution: 0.${\arcsec}$33 & Time \\
 & &  Cadence: 3.6 s & Resolution: 14 s \\
\hline
\hline

 \end{tabular} 
\label{tab:table1}
\end{table} 

\section{Observations} \label{obs}
\subsection{Birth of  the active region}
A  mini-flare (B6.7 X-ray class) and its associated jet was initiated in AR NOAA 12736 located  at N09 W60 on 22 March 2019 around  02:02 UT. The magnetic field evolution and AIA observations of the AR are presented in Sect. \ref{sec:mag} and \ref{sec:AFS} respectively. 

\subsubsection{Magnetic field}
\label{sec:mag}
\begin{figure*}[ht!]
\centering
\includegraphics[width=1.0
\textwidth]{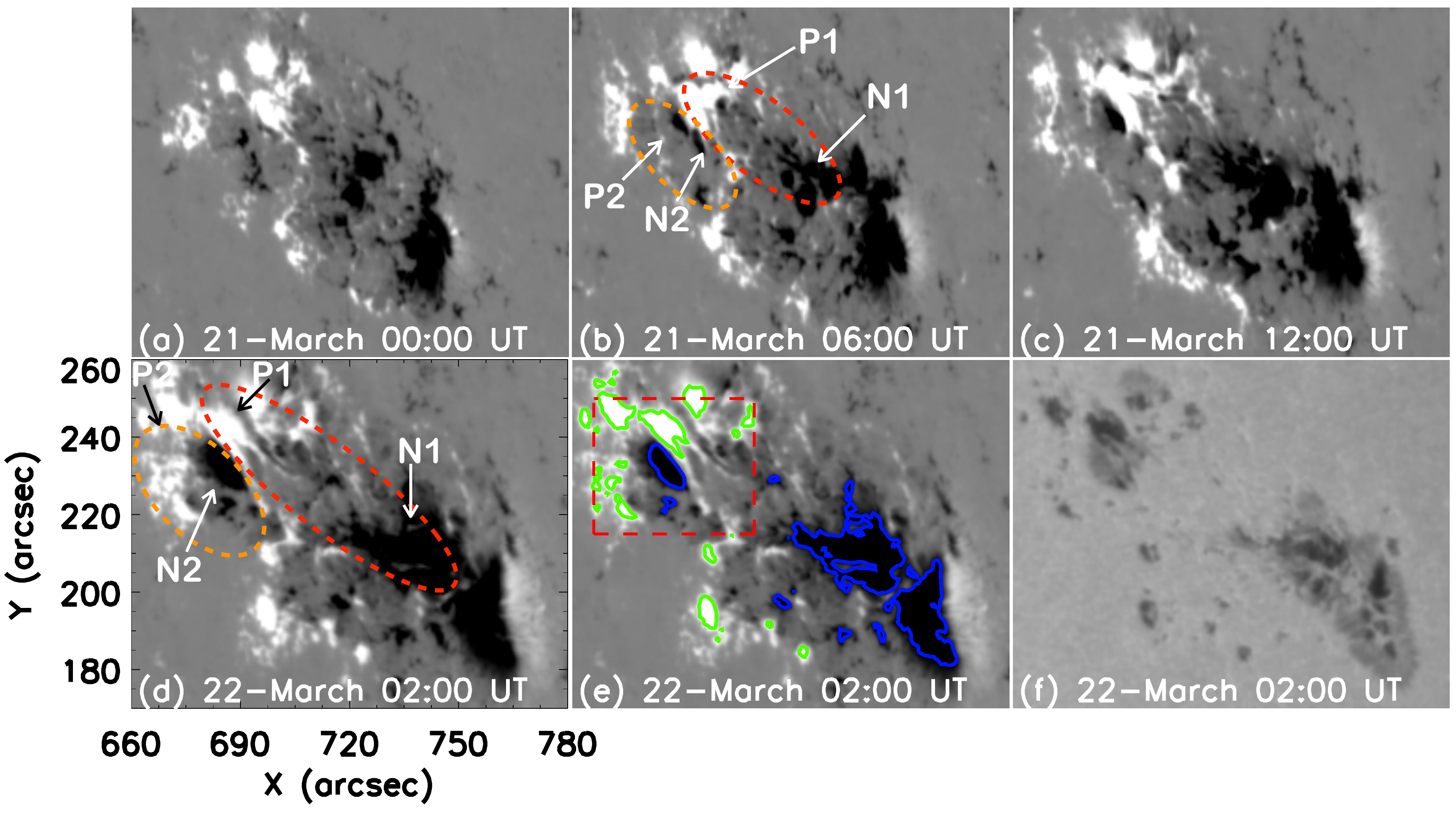}
\caption{Panels (a--e): HMI longitudinal  magnetograms  of AR  NOAA 12736 showing the evolution of the magnetic polarities. The jet reconnection is occurring between the two large emerging flux areas EMF1 (P1, N1) and EMF2 (P2, N2) encompassed in the  two ovals drawn in panels (b) and  (d). The  emergence  of the EMFs is shown  by  the extension of the two  ovals between these two times.  Panel (f):  HMI continuum image showing the sunspots  and pores of the AR.  The blue and green contours are for negative, and positive magnetic field polarities with label $\pm$ 300 Gauss. The red rectangular box in panel (e) shows the FOV of Fig. \ref{flux} (a) and  Fig. \ref{vector} (a) and (b).}
\label{HMI}
\end{figure*}

\begin{figure*}[ht!]
\centering
\includegraphics[width=1.0
\textwidth]{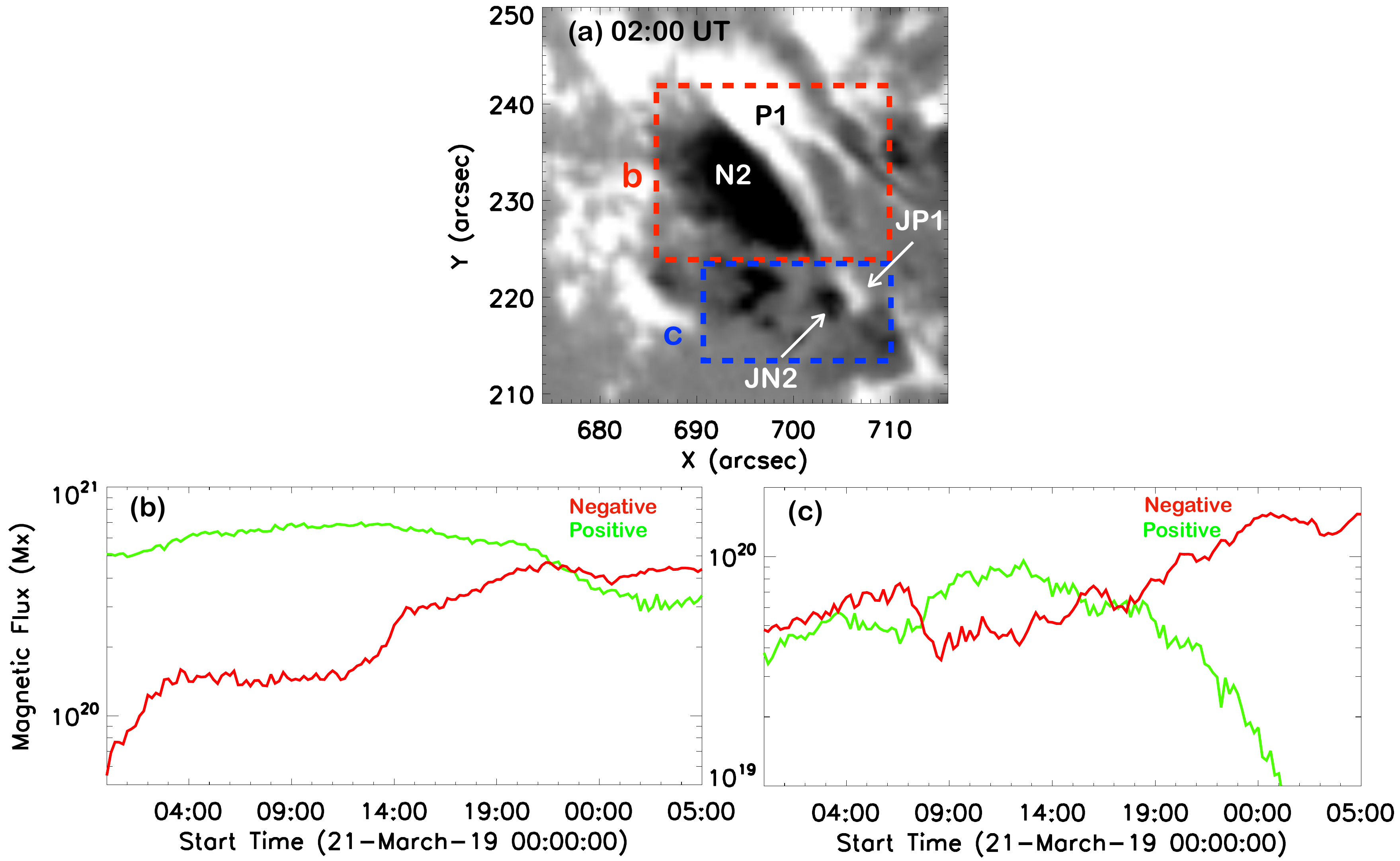}
\caption{Magnetic flux cancellation in two areas including  the major  bipole (P1, N2)   (red box (b))  and the small jet bipole  (JP1, JP2) (blue box (c)), respectively.  In panel (a) a zoom of  the longitudinal magnetic map including the two bipoles is presented (FOV is the red box in  Fig. \ref{HMI} (e)). Panels (b and c): variation of the magnetic flux in the red  and blue boxes.
}
\label{flux}
\end{figure*}

  The AR 12736 was emerging progressively since  19 March 2019. On 21 March, we note two emerging flux regions ({EMFs)} elongated along the  north-east to south-west  direction (see the ovals in Fig. \ref{HMI} (b and d)). The first emerging flux region (EMF1) is the main component of the AR, with negative leading polarity and positive following polarity. The second emerging flux region (EMF2) consists of  many fragmented negative polarities which are travelling very fast as the emerging flux  is  expanding  towards  south  and  squeeze the positive polarity of EMF1.  Consequently, a very high magnetic field  gradient is observed perpendicularly to the polarity inversion line (PIL) between the squeezed polarities: `P1' positive polarity belonging to EMF1 and `N2' negative polarity belonging to EMF2 (see the red box in Fig. \ref{HMI} (e)). The negative polarity N2 is surrounded by positive polarities P1 on the right side and P2 on the left side and top. This topology is classical with the aim of getting a null point, as we see discuss further on in this paper.  In the HMI movie (animation is attached as MOV1), we note the fast sliding motion of  negative polarity N2 towards  the south and the motion of  positive polarity P1 in the opposite direction, which  creates a strong shear between them. Along this PIL, we distinguish that at the time of the flare observations, we principally observe two bipoles emanating from the two EMFs in the diagonal of the box (NE-SW):  a large north bipole (P1, N2) and a very tiny bipole  (JP1, JN2) in the south, which was detached  progressively from the northern N2 polarity a few hours before (the animation is attached as MOV1 and explained in Sect. \ref{reconnection}).  We computed the flux budget for these two bipoles and we found a significant decrease of the positive flux in the two  boxes, each of them including a bipole (Fig. \ref{flux}). We interpret these decreases by magnetic cancelling flux. The tiny bipole is labeled  with `J' like ``jet" because this is the
 the location where the jet took place.

 \subsubsection{Arch filament systems (AFS)}
 \label{sec:AFS}
\begin{figure*}
\includegraphics[width=1.0
\textwidth]{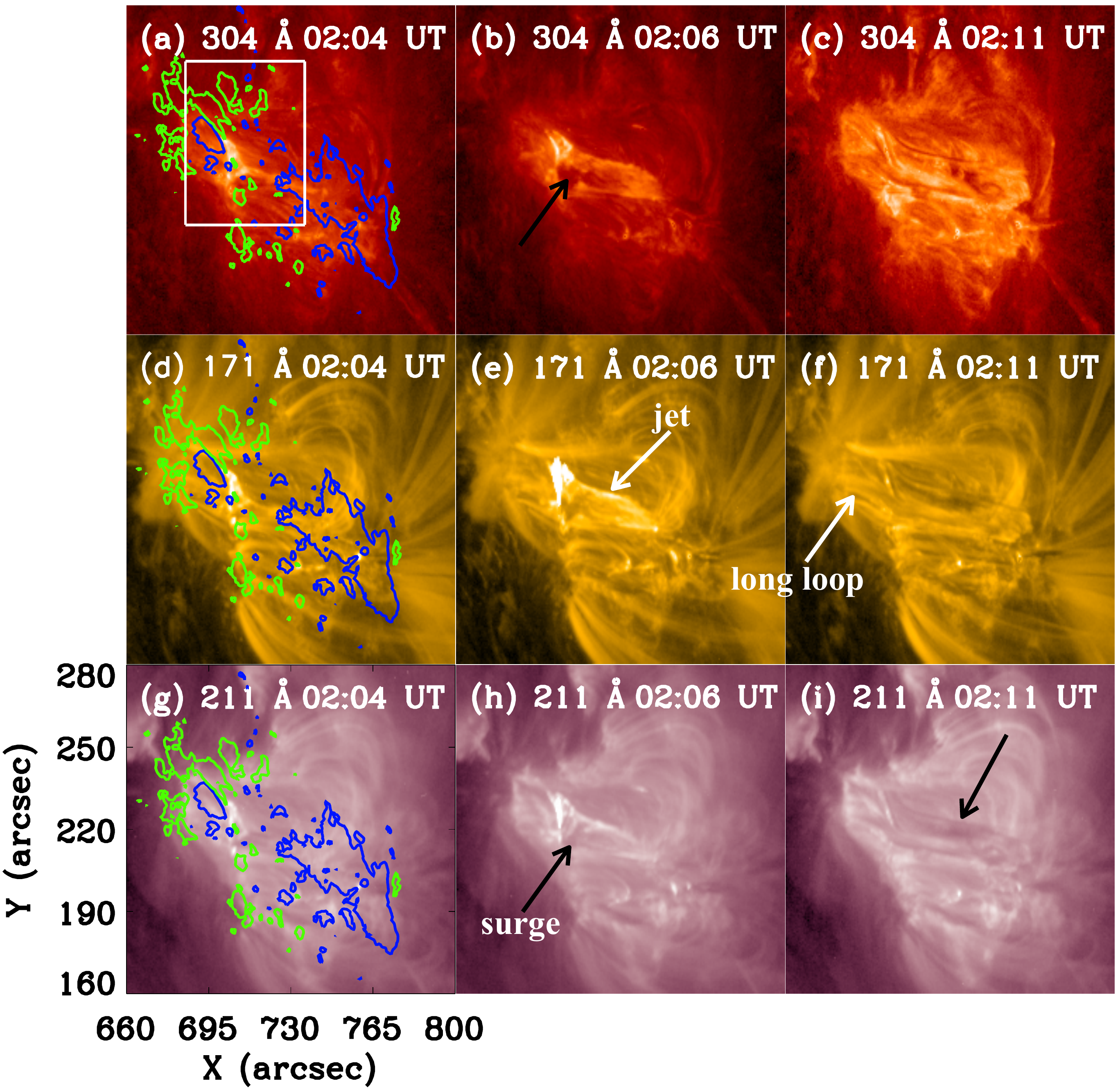}
\caption{Solar jet  and surge observed in different AIA/EUV channels (304 \AA, 171 \AA, and 211 \AA) on March 22 2019 between 02:04 UT and 02:11 UT  in NOAA AR 12736. The black arrows in (b) and (h) indicate a dark area corresponding to a surge, the white arrow points the jet in (e)  and   a long loop created after the reconnection in (f), the black arrow in (i)  an AFS. In the first column the images are overlaid with HMI longitudinal magnetic field contours ($\pm$ 300 Gauss) for positive and negative magnetic polarities with green and blue colours, respectively. The white box in panel (a) indicates the FOV of Fig. \ref{AIA_IRIS}. The small bipole center is at 710$\arcsec$, 220$\arcsec$, and the major bipole center is at 695$\arcsec$,  230$\arcsec$.
}
\label{AIA1}
\end{figure*}

  The AIA observations cover the  AR and the full development of the jet in  multi-temperatures provided by all the sample of  AIA filters from 304 \AA\ to 94 \AA, all along the range of temperatures from 10$^5$ K to 10$^7$ K (Fig. \ref{AIA1} and Fig. \ref{AIA2} along with the related movies (animations are attached as MOV2a-d in AIA 131 \AA, 171 \AA, 211 \AA, and 304 \AA).  Contours  of  longitudinal magnetic fields ($\pm$ 300 Gauss) are overlaid on the AIA images to specify the location of the small bipole JP1-JN2 at the jet base.
 AFS are well visible over the  two emerging flux EMF1 and EMF2 with cool and hot low lying loops joining the positive  and negative polarities for each of them, P1 and N1 on the  west side and  P2 and N2 on the east side.  
 Filaments  belonging to these AFS are 
 particularly  visible as dark structures due to absorption mechanism in 171 \AA,   131 \AA,  and 193 \AA\ at 02:06 UT (Fig. \ref{AIA1} panel e and Fig. \ref{AIA2} panels b, e). These filaments are parallel to each other, oriented more or less NE to SW
 from P1 to N1 and from P2 to N2 in the direction of the extension of each  EMF.   
  They do not lie along any PIL and, therefore, they do not correspond to the usual definition of filaments; rather, they  are  more or less  perpendicular to the PIL in each EMF.  Filters AIA 171 \AA\, and 193 \AA\, are good proxies for detecting cool structures visible in H$\alpha$. 
  At these wavelengths, the EUV emission is absorbed by the  hydrogen and helium continua  \citep{Anzer2005}. The opacity of the hydrogen and helium resonance continua at 171 \AA\, is almost two orders of magnitude lower than the  Lyman continuum  opacity  at 912 \AA\, 
  and thus similar to the H$\alpha$ line opacity \citep{Schmieder2004}. 
  We confirm the presence of H$\alpha$ filaments/AFS by looking at the  H$\alpha$ images (Figure \ref{NVST}). The two AFS over the two EMFs are well identified.  On the west side the AFS are
  dense and long, with some  narrow arch filaments overlying the bright corridor of the PIL (N2-P1)  and the dome of EMF2 before the burst  (panel b). The AFS over EMF2 on the east side have a fan structure with an anchorage all around the negative polarities N2  and the other in the positive polarities P1 and P2. It gives the impression of an `anemone' structure which is frequently observed for jets  triggered  by emerging flux  \citep{Shibata1982, Schmieder2013, Joshi2020}.
In order to follow the jet development with AIA,  we focus on the FOV covering 
  the two bipoles identified in the previous section. For more details, see the white box in Fig. \ref{AIA1} (a).

 \subsection{Observations of the jet} 
 \subsubsection{Jets in AIA images}
\label{jet_AIA} 
\begin{figure*}
\includegraphics[width=1.0
\textwidth]{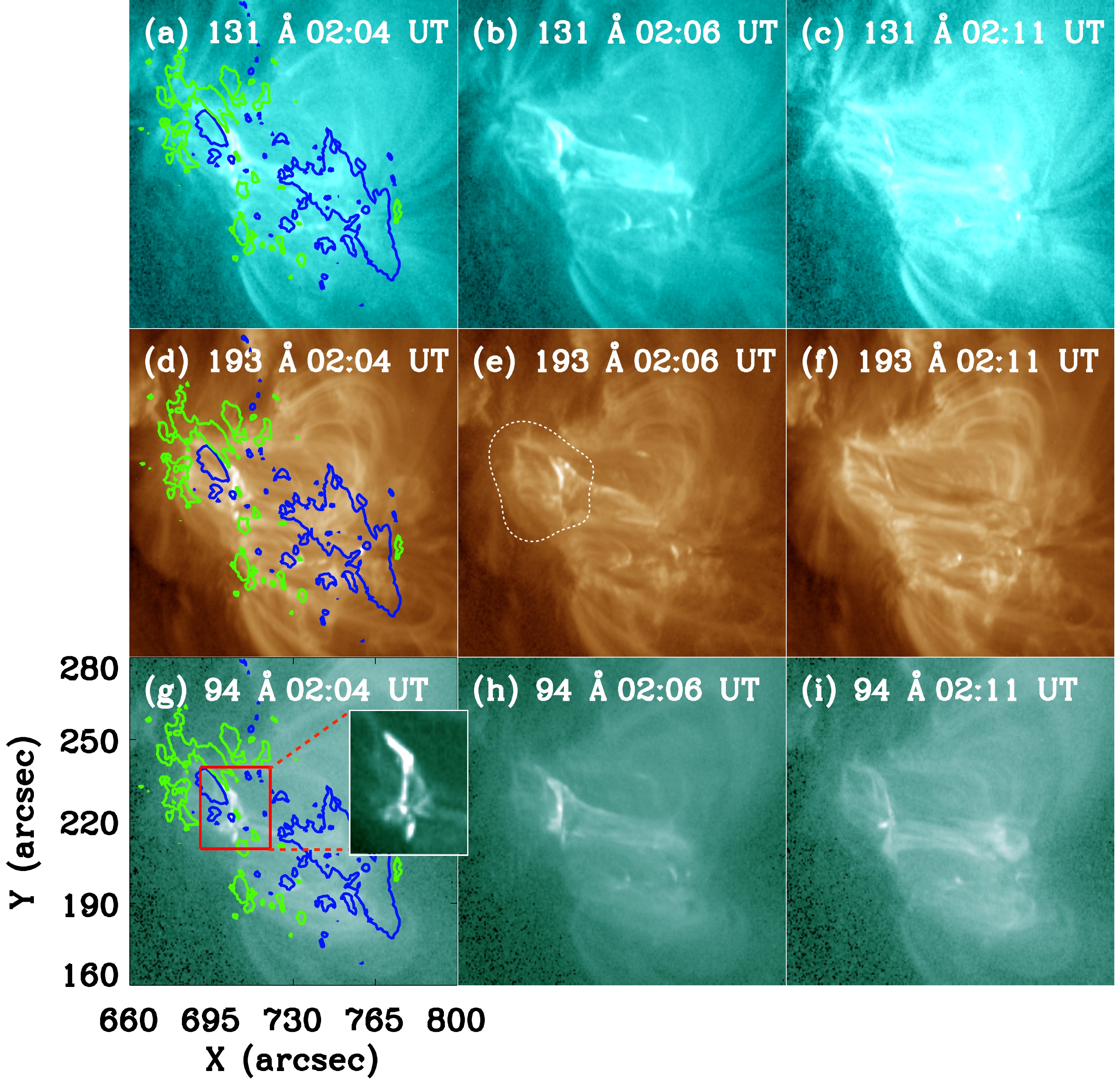}
\caption{{ Solar jet  and surge observed in the other AIA/EUV channels (131 \AA, 193 \AA and 94 \AA). The first column shows the images overlaid with HMI longitudinal magnetic field contours ($\pm$ 300 Gauss) for positive and negative magnetic polarities with green and blue colours, respectively. Panel (e) shows a dome structure at the jet base highlighted by a white dashed contour.
In panel (g), the red box shows a zoom view of the brightening at the jet base overlaid by the contours of   the small bipole (710$\arcsec$, 220$\arcsec$) and the large bipole (695$\arcsec$,  230$\arcsec$).}}
\label{AIA2}
\end{figure*}

 It is interesting to see that activity had started before the onset of the jet, with very bright north-south tiny   threads observed above the PIL between the two EMFs and, more precisely, between the part of the  PIL in the northern bipole (P1-N2)  and continuing into the south tiny bipole (JP1-JN2)  around 02:01 UT until 02:04 UT. See inserted zoom image in Fig. \ref{AIA2} (g)). 
 ?

\begin{figure*}
\includegraphics[width=1.0
\textwidth]{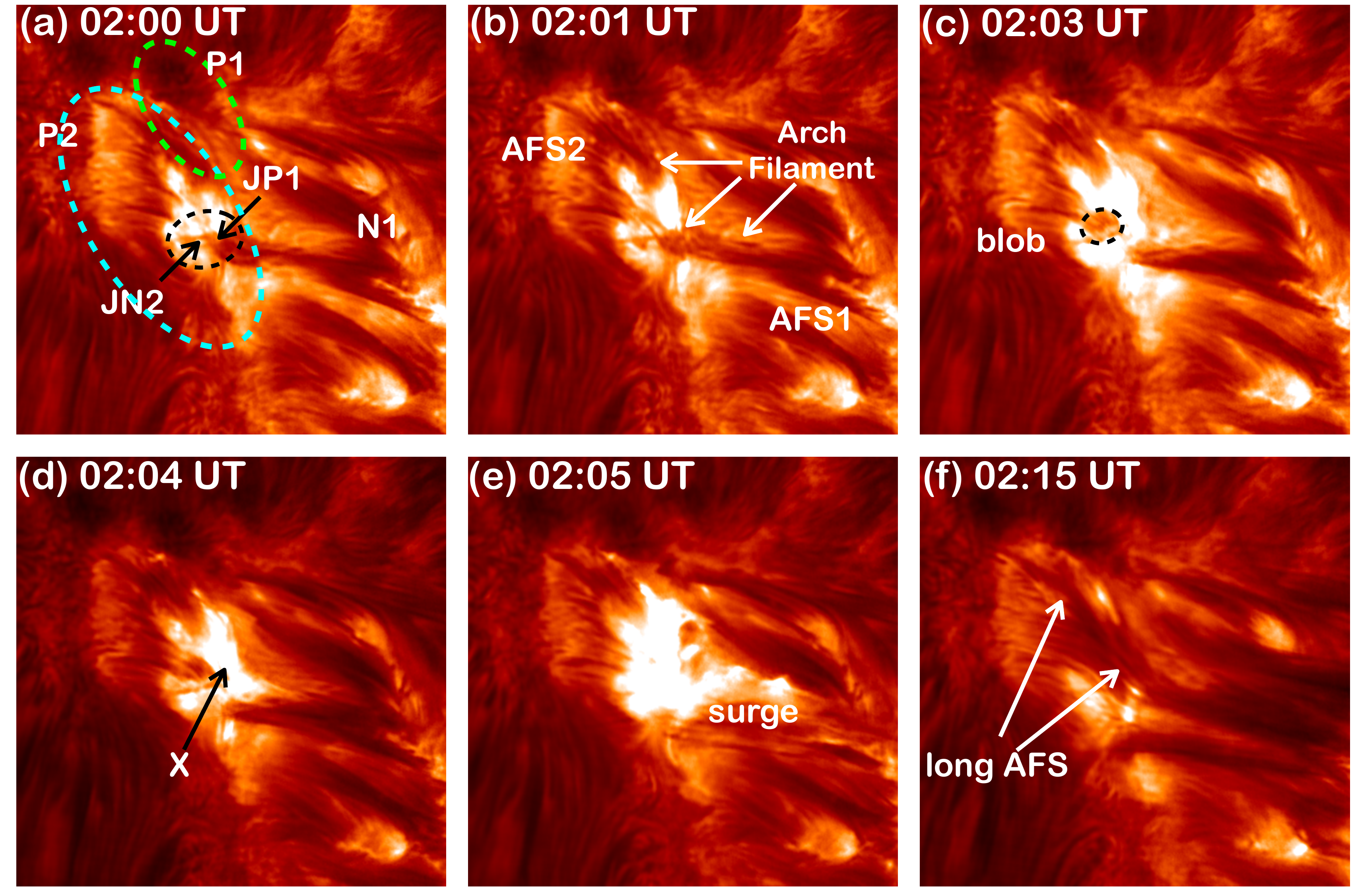}
\caption{H$\alpha$ line center observations of the AR NOAA 12736 with the NVST telescope
for four times before (a-d) and one after (f) the surge extension. In panel (a), the cyan oval encircles the emerging flux EMF2 like in Figure \ref{HMI} which corresponds to a dome (see Sect. \ref{jet_AIA}), the green circle indicates the bipole P1-N2, the black circle the bipole JP1-JN2. In panel (b), we highlight the two systems of AFS (AFS1 and AFS2), with the white arrows indicating long AFS overlying both AFS1 and AFS2 structures. In panel (c), we see a  dark blob in the middle of the brightening. In panel (d) the `X' point is indicated where the reconnection occurs. In panel (e), the white, more or less horizontal  material represents the beginning of the surge ejection with some knots. In panel (f), we again see the long AFS. We have accurately rotated the data provided by the NVST team to make them comparable with the AIA and IRIS observations.
The image size is 65$\arcsec$ $\times$ 65$\arcsec$.
}

\label{NVST}
\end{figure*}

The bright signature around  a  dome structure overlying the EMF2 at the jet base is highlighted by a white dashed contour around it, as seen in Fig. \ref{AIA2} (e). This dome is highlighted by the small fibrils with an  asymmetrical anemone shape visible in the NVST images (Figure \ref{NVST} b). Along the west side of the dome, the brightening with an north-south arch-shape is visible 
 in all AIA channels
 before the  burst  indicates already the presence of  hot plasma (T between 10$^4$ to 10$^6$ K). This region between the two EMFs corresponds to   quasi-separatrix layers (QSLs) \citep{Demoulin1996}, where  a strong high electric current  develops and heats the plasma, as shown here  by the arch-shape brightening. This is subsequently confirmed with the analysis of the photospheric vector magnetic field  maps in Sect. \ref{reconnection}. These QSLs have been calculated in \citet{Yang2020} and are well-identified in this region. QSLs are robust structures but their  localisation is not commonly defined with any substantial accuracy \citep{Dalmasse2015,Joshi2019}. In our case, the moving polarities is a problem for the exact localisation  of QSLs.
  
At the same time (02:04 UT), a jet with two branches inserting a surge is also observed.
  Dark absorbing material is visible 
   at $\approx$ 02:02 UT, resembling a blob with no really  defined shape in the southern part of the arch-shaped brightening (see Figure \ref{NVST} panel c). It then extends to the north and, finally, goes along the jet direction. The surge appears as a dark area in the images because of  the absorption of the UV emission. Therefore, the dark part observed in 171 \AA\, and in the other AIA filters, that is, (211, 193, 94 \AA) should  correspond to cool plasma as seen in H$\alpha$  as we mentioned in the previous section \citep{Schmieder2004}. It is why 
   we call it a `surge' (Fig. \ref{AIA1} and Fig. \ref{AIA2}). The surge appears as a bright structure in the NVST images (Figure \ref{NVST} e). However,   the surge   consists of  cool plasma because H$\alpha$ formation  temperature is lower than 1.5 $\times$ 10$^4$ K.

  The jet base  on the east  side  of the EMF1  is extended  along a more or less  north-south  direction, along
    the PIL between EMF1 and EMF2 (P1 and N2; see Fig. \ref{AIA2} g) and the  jet top  on the west  side  of the EMF1 is limited at the location of  
    N1.
  At 02:11 UT  and  over a few minutes up until 02:18 UT, we can observe long  bright and dark AFS striding over the two EMFs (see right columns of Fig. \ref{AIA1} and Fig. \ref{AIA2}). 
  The characteristics of the jet are the following: length around 50 Mm, the base width between 15-20 Mm. The jet lifetime is 
  between 02:02 UT to 02:11 UT.

  In  AIA, a 304 \AA\ movie  starting at 20:00 UT on 21 March 2019 and running until 22 March 2019 at 05:00 UT, we made several observations of a mini-flare, approximately at the same location, nearly one every hour and generally not accompanied by such a wide jet. The detail of the recurrent mini-flares is as follows: at the beginning of the movie,  a  mini-flare is visible at 20:00 UT, then  at  20:27 UT, 21:28 UT, 22:51 on 21 March and  at 00:39 UT, 01:25 UT,  01:39 UT, 02:03 UT on 22 March. Regularly, before each  burst, we can clearly see  two AFSs:  one over EMF1 in the west and one  over EMF2 in the east. After the burst, long-arch filaments connect the extreme eastern polarity P2 to the extreme western polarity N1. Prior to our mini-flare and jet (around 01:59 UT), the two AFSs were separated by an area with mixed bright and dark patches. Then at 02:09 UT, there is a long system of arch filaments. At 02:28 UT,  
   when the phase of the activity is over, the initial  configuration with the two distinguished AFSs, just as before the jet, is restored. This chain of mini-flare and ejection is recurrent.
 
 \subsubsection{Transverse projected  velocity}
 
\begin{figure*}[ht!]
\centering
\includegraphics[width=1.0
\textwidth]{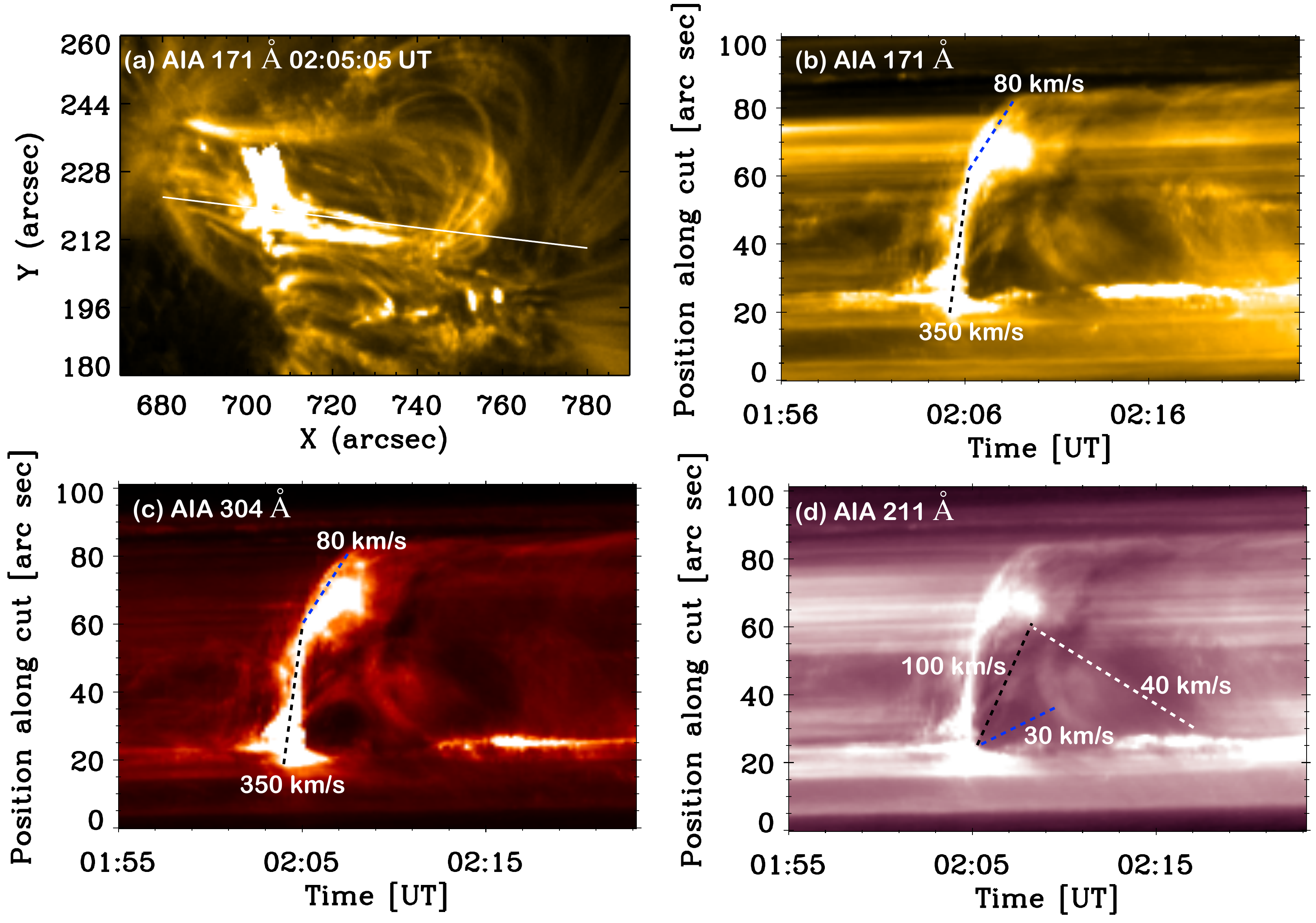}
\caption{Height-time profile for the jet in different AIA wavelengths (b-d). The location of the slit (width = 8 pixels) along the plasma flow direction, which is used to generate the height-time plot, is shown as the white solid line in panel (a).
The jet and surge both showed a two phase eruption, i.e. fast and slow. The jet is erupted with a high speed of $\approx$ 350 km s$^{-1}$ and after that it is slowed down to
$\approx$ 80 km s$^{-1}$ (panel (b-c)). The surge is erupted with a maximum speed of $\approx$ 100 km s$^{-1}$ and the dense material came back to the source region with a speed of $\approx$ 40 km s$^{-1}$ (panel (d)).
\label{timeslice1}}
\end{figure*}

  To analyse the kinematics of the observed jet, we created the projected height-time profiles of the jet in different AIA wavebands (171 \AA, 211 \AA, and 304 \AA), which is presented in Fig. \ref{timeslice1}. To obtain this height-time plot, we chose a broad slit (width = 8 pixels) to cover the plasma outflow. 
 The average jet speed along the slit direction shows two different slopes: a steep slope in the starting phase (till 02:05 UT) of the jet eruption with an average speed of 350 km s$^{-1}$ and a slow phase of 80 km s$^{-1}$ in the later stage from 02:05 UT to 02:08 UT.
 The slow phase may be due to the presence of loop system in the path of the jet. It seems that when the jet material is passed through this loop system, it decelerated and,
 finally, it stopped.

 The cool material visible as an absorbing feature in AIA channels is detected about one minute after the hot jet with no well defined speed. The cool material  appears to be present along the line of sight in small patches but it is not moving towards the west as the jet is doing. Later, the cool material (or surge) is escaping in  slightly different directions than the jet. Among it, we could identify some blobs with different projected speeds
 ($\approx$ 100 and 30 km s$^{-1}$). Then the cool material  came back at a speed of $\approx$ - 40 km s$^{-1}$ (Fig \ref{timeslice1} (d)). According to the location of the AR (W60), these velocities are underestimated by a factor of cos(60). The positive velocities correspond to the material that is heading away from the observer's view.

  \subsection{Corresponding structures in IRIS SJIs and AIA 304 \AA}
  \label{comparison}
\begin{figure*}[ht!]
\centering
\includegraphics[width=1.0
\textwidth]{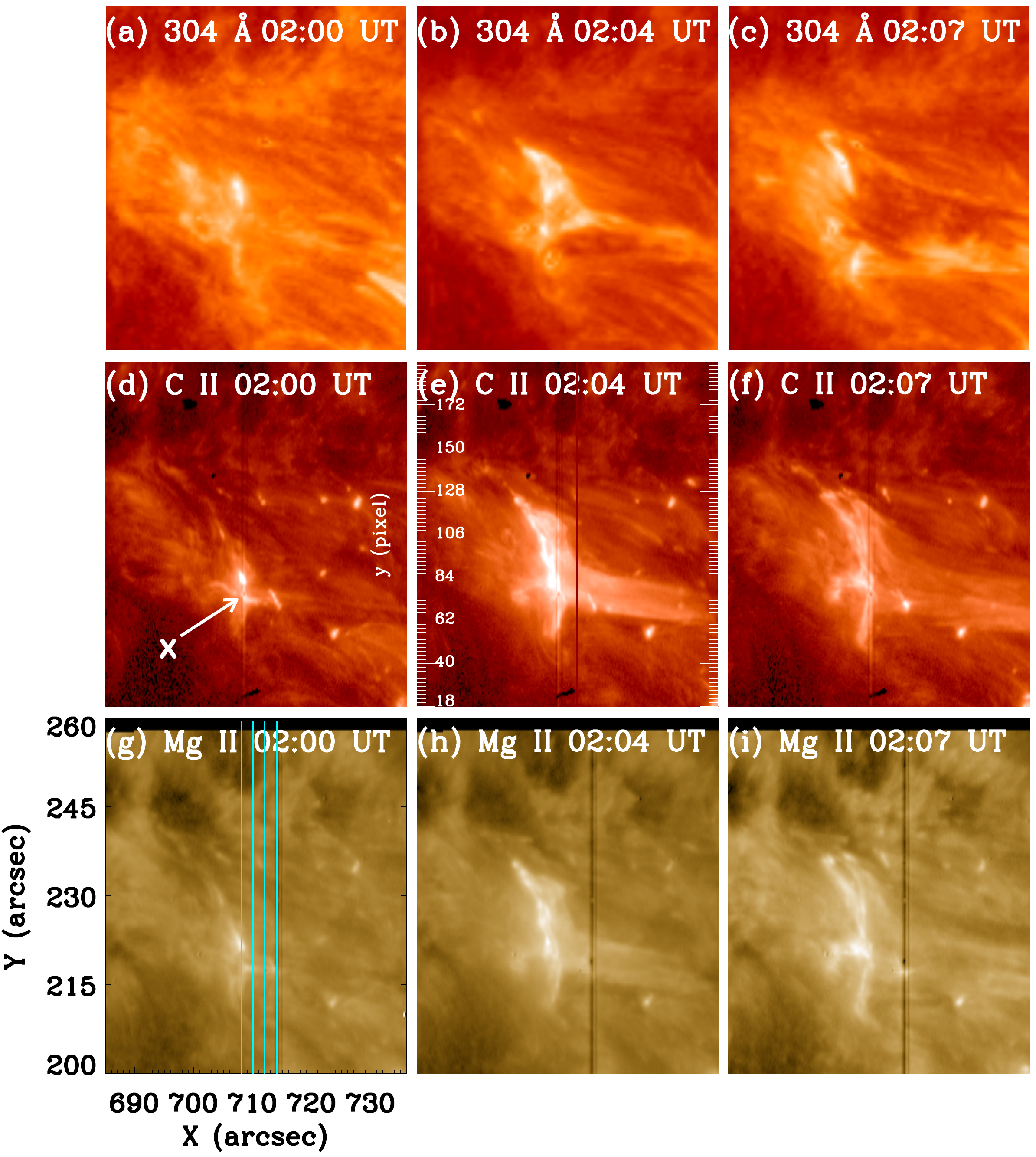}
\caption{Solar jet observed in  co-aligned  images of EUV AIA 304 \AA\ (top), of IRIS C II  SJI (middle), and of IRIS Mg II SJI  (bottom). The  four positions of the slit in the raster mode  are shown with vertical cyan lines in panel (g). The dark vertical line in each panel (d-i) is   the footprint of  one position of the slit during the SJI shooting. { In IRIS CII  SJI images  (middle row), the reconnection point is  indicated
  as 'X' in panel  (d)  and 
 the original observed pixel values along the  y axis   are displayed in panel (e).
The FOV of these panels is indicated in Fig. \ref{AIA1} (a)}. In this figure we keep the nominal coordinates of IRIS which are translated by + 4 arcsec in x and +3 in y compared to the  AIA coordinates. 
\label{AIA_IRIS}}
\end{figure*}

  The  analysis of IRIS data,  all along the evolution of the jet, shows a good correspondence between the structures visible in AIA 304 \AA\  and in IRIS C\footnotesize{II} \normalsize SJIs (animations are attached as MOV2d (AIA 304 \AA), and MOV3 (IRIS 1330 \AA)). This correspondence is  summarised in Fig. \ref{AIA_IRIS}. We note that the nominal coordinates of IRIS  in the file headers do not correspond to the nominal coordinates of AIA. Therefore, we had to shift  the FOV of
AIA by 4 arcsec in x-axis and 3 arcsec in y-axis to obtain a good co-alignment.
  The IRIS slit,  with its four positions, crosses the bright zone corresponding to the jet base, namely, the dome top, which is supposed to be the reconnection site  along a few pixels between around pixels 60 to 120  in the left slit position (Fig. \ref{AIA_IRIS} (e)).

 Around 02:00 UT, in the 304 \AA\, image  as well as in C\footnotesize{II}  \normalsize and Mg \footnotesize{II} \normalsize IRIS  SJIs,  small bright threads along two vertical paths that are mixed with 
 tiny round-shape  darker areas are visible in the middle of the FOV  where the reconnection  occurred  (Fig. \ref{AIA_IRIS} a, d, and g). It is clear that in this small zone, there is no 
north-south  filament  along the PIL (N2-P1)   which would be visible by absorption in AIA 193 \AA.  The very  light-dark  filament-type structure with a vague sigmoidal shape in the NVST images that is localised at this place is, in fact, part of the AFS  (see Figure \ref{NVST} b) because there is no sigmoid  visible in the hot channels of AIA, where plasma should be heated   due to high electric currents along a sigmoid \citep{Barczynski2020}. In the north of this zone,  long-lying,   more or less east-west  AFS, as well as, on both sides of the zone (pixels 60-120), the short AFS-overlying EMF1 and EMF2 are visible. The short AFS-overlying EMF2  have  a dome  shape like the asymmetrical  anemone formed by the fibrils visible
in the NVST images (Figure \ref{NVST}).

 The location of the onset of the  mini-flare
 is indicated by  the point `X'  at the crossing location between the  arch-shape QSL and an east-west bright  line in Fig. \ref{AIA_IRIS} (d)). The location of the `X' point   in IRIS observation   is at 709$\arcsec$, 218$\arcsec$ and  in AIA it  is at 705$\arcsec$, 215$\arcsec$. In  Figure   \ref{AIA_IRIS} (top panels), we translated the AIA images to obtain a good co-alignment with IRIS images.
 Around 02:03 - 02:04 UT, the arch-shape  QSL  was brightening  and the flare started 
 with  the onset of the jet  ejection visible  in  the C\footnotesize{II} \normalsize and Mg\footnotesize{II} \normalsize SJIs (see MOV3 in C\footnotesize{II} \normalsize and Fig. \ref{AIA_IRIS} (b,e))). The bright jet was obscured by dark material in front of it, which is the surge; both the jet and the surge were extending at the same time around 02:04 UT.
 From 02:05 UT to 02:07 UT, the jet  extended along  two bright branches with  a dark area between them (Fig. \ref{AIA_IRIS} ((c),(f)) and Fig. \ref{tilt} (a)). 
 At 02:07 UT, AIA 304 \AA\ image shows the extension of the  surge covering all the bright jet. The dark area is  due to the absorption of the 304 \AA\,  emission by He continua \citep{Anzer2005} (Fig. \ref{AIA_IRIS} (c)). In  the C\footnotesize{II} \normalsize and Mg\footnotesize{II} \normalsize filters, it is not so pronounced because of the large band-pass of IRIS filters relative to the width of the lines  and the low emission of the lines in the jets (Fig. \ref{AIA_IRIS} (f),(i)). Nevertheless, we  can still distinguish a bright EW-elongated jet in the south  and  some dark area above it that might correspond to the surge. This is confirmed in the NVST images (Figure \ref{NVST}).
 
\subsection{Tilt of IRIS spectra}

\begin{figure*}[ht!]
\centering
\includegraphics[width=0.90
\textwidth]{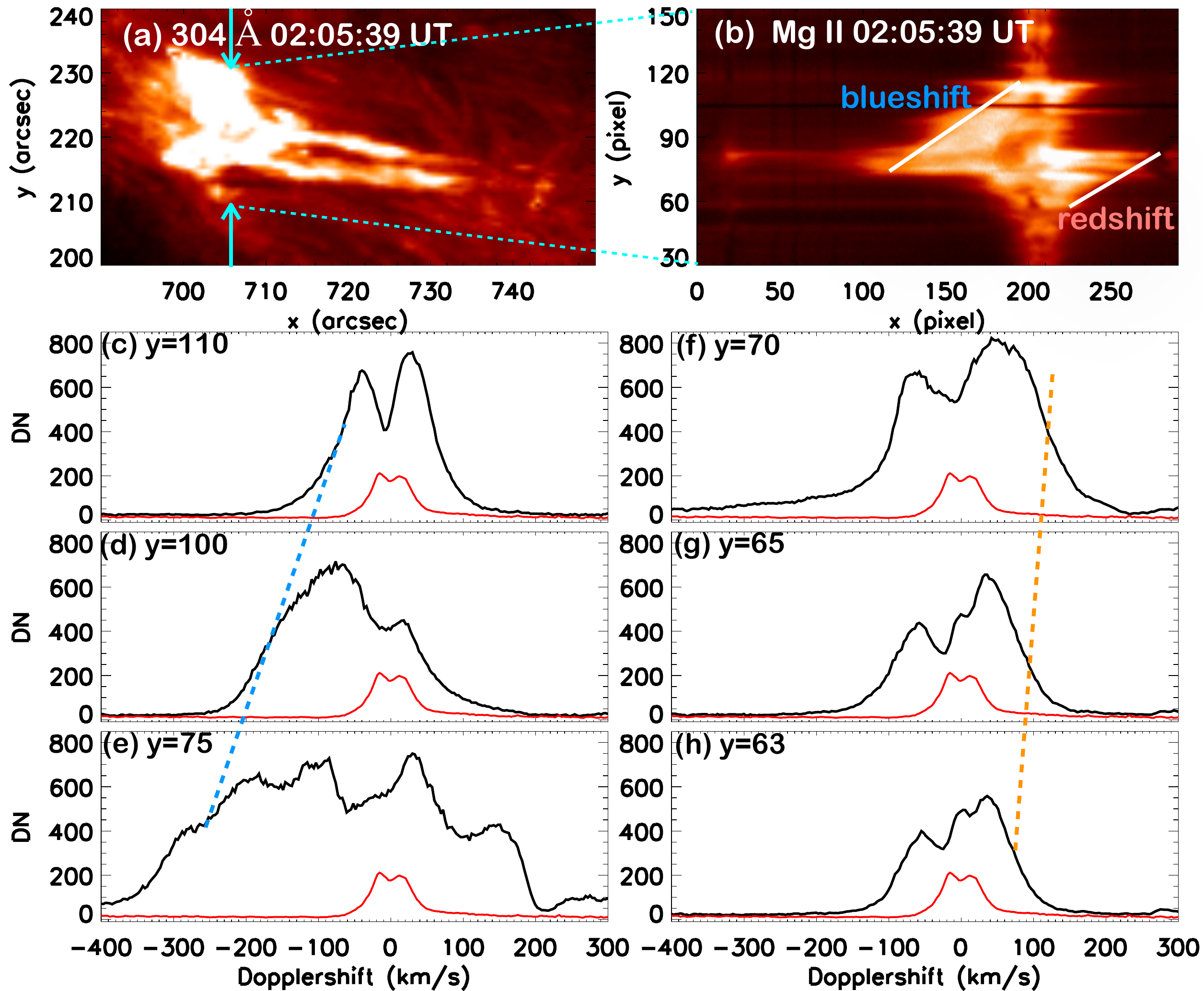}
\caption{Mini-flare and the  bright jet  with two branches  inserting a cool dense surge in AIA 304 \AA\, (panel a). The cyan arrows show the location of the slit position 1. The Mg II k line spectra corresponding to the slit position 1 which is crossing the reconnection site is  shown in panel (b).
The red and blue shift  wings
are shown by the white tilted lines on the left (blueshift) and right (redshift) in the spectra. The bottom rows (c-h) show the Mg II k line profiles for  different y values using unit of velocity in the x axis.
Panels c-e concern  the blue shift  profiles.   shown by  the tilted blue dashed line corresponding to  strong blueshifts (-300 to -100 km s$^{-1}$). Panels f-h show the red shift profiles. Redshifts (80 to 100 km s$^{-1}$)  are  shown by the red dashed tilted line. The red and blue dashed  lines are passing through the inflexion points of the line profiles in panels c-h).}
\label{tilt}
\end{figure*}

 Over the course of the reconnection phase (starting at 02:04 UT) the Mg\footnotesize{II} \normalsize , C\footnotesize{II,} \normalsize and Si\footnotesize{IV} \normalsize spectra show extended blue and red wings around the pixel value of 80. As an example, we show  Mg\footnotesize{II k} \normalsize line spectra (Fig. \ref{tilt} (b)).  
 At the pixel value 80, the  Mg\footnotesize{II} \normalsize line profile is the most extended one on the blue and red sides like in bidirectional outflows. This kind of bidirectional flow has been interpreted as being the site of reconnection in some events \citep{Ruan2019}. Therefore,  we consider  this zone (around the pixel value 80) to be  the reconnection site.

\begin{figure*}[ht!]
\centering
\includegraphics[width=0.90
\textwidth]{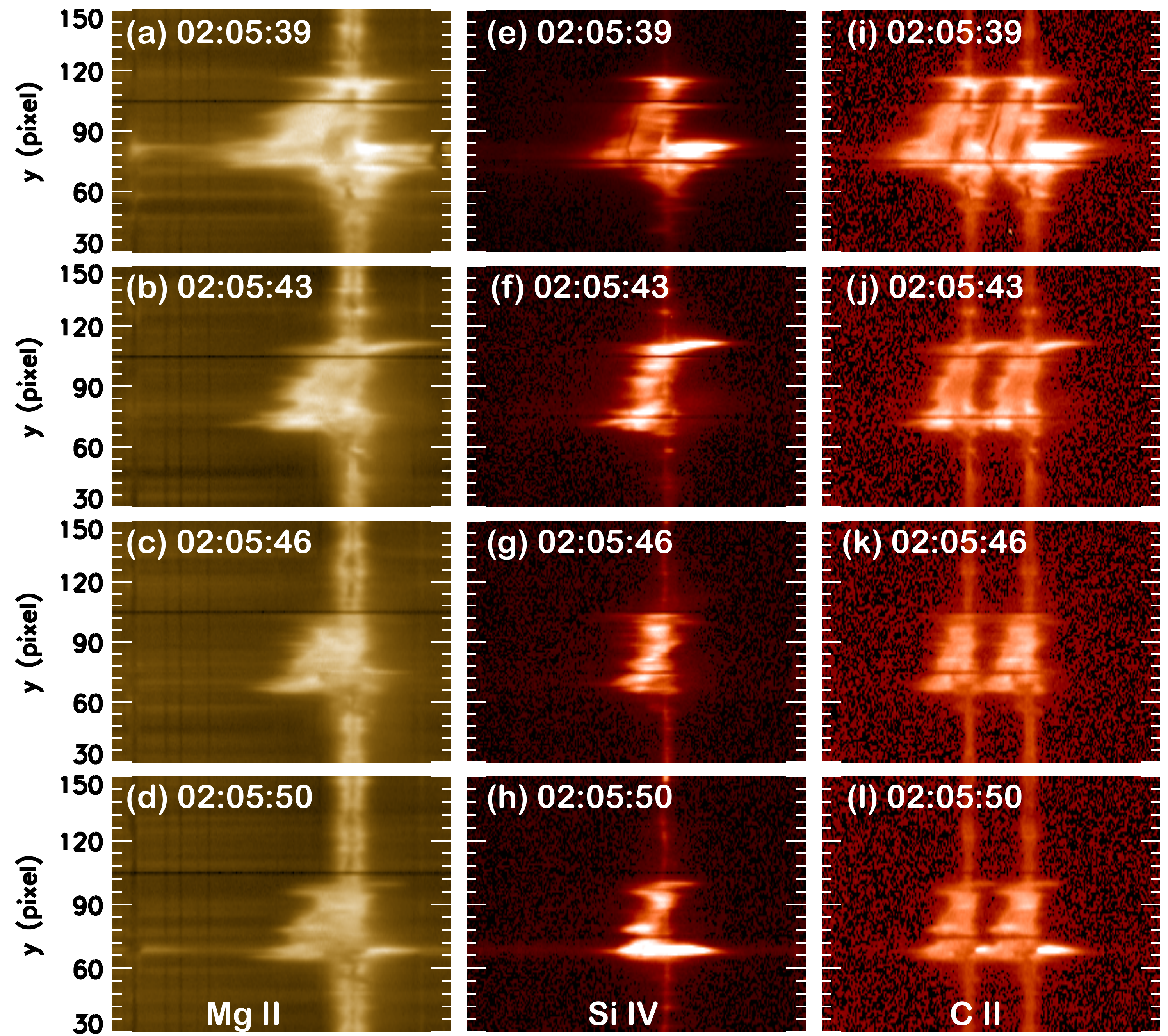}
\caption{Tilt observed in the three lines Mg II (left column), C II (middle column), and Si IV (right column) observed with IRIS instrument. The four different rows are showing the spectras at the  four different slit positions  from east to west: 1, 2, 3, and 4.}
\label{3_tilt}
\end{figure*}

 Rapidly (in less than one minute), we see an extension of the brightening  of the wings of Mg\footnotesize{II k} \normalsize spectra in pixels along the slit in the central zone (pixel value 60 to pixel value 120) (Fig. \ref{tilt} (a)).
 In the north and south parts of the reconnection site, we note that the spectra shows a tilt. The Mg II profiles are not symmetrical all along the slits;  they present some bilateral flows only in a few pixels around y = 79. Otherwise, they have extended blue wings for example for pixel y $>$ 85  and not corresponding red wings. It is the reason that we do not consider that the bilateral flows exist all along the slit (10 Mm long). Therefore,  the  existence of this gradient and tilt is obvious. 
   The tilt is characterised by the gradient of the Dopplershifts that exist for profiles along the slit at a given  time.
The line profiles of  Mg\footnotesize{II k} \normalsize line
  show  important  extensions of the wings   at 02:05:39 UT (Fig. \ref{tilt} (c-h)). 
They     predominantly show an  extended blue wing in the northern
part (pixel value 75 to pixel value 110 (c-e)) with decreasing blueshifts 
at y = 110; they are roughly symmetric
in the middle of the brightening (y = 80) with  more extended blue wings nevertheless (until -300 km s$^{-1}$).
In the southern part of the brightening, the profiles have a dominant
 enhancement in the  red wing 
(pixel value 63 to pixel value 70 (f-h)).
The x-axis of the Fig. \ref{tilt} shows  Dopplershifts in km s$^{-1}$.  
These Dopplershifts do not really correspond to up and down flows because  the region is located at 60${^\circ}$ in the west.  Therefore, the blue-shifted material is going, in fact, to the left of the reconnection site over the EMF2 and not in the direction of the jet. This means that all  cool material visible at -300 km s$^{-1}$ for which the  emission is relatively high in the Mg\footnotesize{II} \normalsize wings is going to the east and the redshifted material is expelled toward the west side as is the jet, with a maximum velocity of 80 km s$^{-1}$. The transverse velocity of the cool material along the west side  has dispersed values of between 30 km s$^{-1}$ and 100 km s$^{-1}$.  This means that one part of the cool red-shifted material could be  nearly normal to the solar surface while the other part would be inclined like  the jet.
A similar behaviour is observed in the four positions of the slit for Mg II, C II, and Si IV lines (Figure \ref{3_tilt}). The tilt in  the four  Si IV spectra is even more readily visible because Si IV is a transition  region line   with only one  emission peak when compared to chromospheric  lines with two peaks.

This  type of tilt spectra along a slit was first observed for prominences \citep{Rompolt1975} and interpreted as rotating prominences before eruption. Thanks to the Solar Ultraviolet Measurements of Emitted Radiation (SUMER) spectrograph onboard SOHO, and now also with the IRIS spectrograph, such a tilt behaviour in the spectra is frequently observed. They are well-known and typically associated
with twist \citep{Pontieu2014}  or rotation \citep{Curdt2012},
or flows of  plasma in helical
structures \citep{Li2014}.  In \citet{Li2014},  the 
long filament crossed by the  IRIS slits changed the direction of its rotation in the middle of the filament. In our observations,  the jet  is rotating   in the same direction in all four positions. The  slit  scanned only 6 arc sec of the jet, mainly capturing the jet base with the dome shape.

 The tilt in our spectra finally reach a length around 60 pixels, which represents around 15 Mm. We interpret this tilt by the rotation of a structure crossed by the slit, the structure being the base of the jet or, possibly, cool plasma that follows helical structures.
The profiles of the Mg\footnotesize{II k} \normalsize line with extended wings resemble the profiles of the IRIS bombs (IBs) that were discovered by \citet{Peter2014}  and analysed by \citet{Grubecka2016} and \citet{Zhao2017}. \citet{Grubecka2016} found that the IBs were formed in the very low atmosphere between 50 to 900 km in the chromosphere. The magnetic configuration of the reconnection site is similar to that of the Ellerman bombs (EBs) in BP regions, where there is no vertical magnetic field \citep{Georgoulis2002,Zhao2017}. We conjecture that between the two EMFs in the QSL region, there is a BP reconnection region as in IBs \citep{Zhao2017}.  The  BP topology in the  region  of the present jet is confirmed in the topological analysis (Sect. \ref{reconnection}).
The cool material which is expelled towards the east could correspond to the dark blobs that previously existed in this area, which were trapped in the BP region while the BP was forming between   two mini-flare events.

\section{Magnetic topology  of the jet region} \label{reconnection}
In Sect. \ref{sec:mag}, we  follow the birth of the AR using HMI longitudinal magnetic field. Here, we analyse 
the magnetic topology of the AR  using the full vector magnetic field to understand the orientation of the magnetic field  lines inside the two bipoles (P1-N2 and JP1-JN2) 
involved in the mini-flare and the jet to confirm the existence of a BP.
\subsection{HMI Magnetic field vector maps}
\label{sec_vector}

The HMI SHARP longitudinal magnetic field movie, with its high cadence, shows
the fast evolution of the EMF2. The negative polarities N2-JN2 were continuously sliding along the positive polarity P1 of EMF1 and  initiating bright points from time to time. Therefore, we used  the  HMI vector magnetic field map at  the closest time of the reconnection at 02:00 UT.
 
\begin{figure*}[ht!]
\centering
\includegraphics[width=1.00
\textwidth]{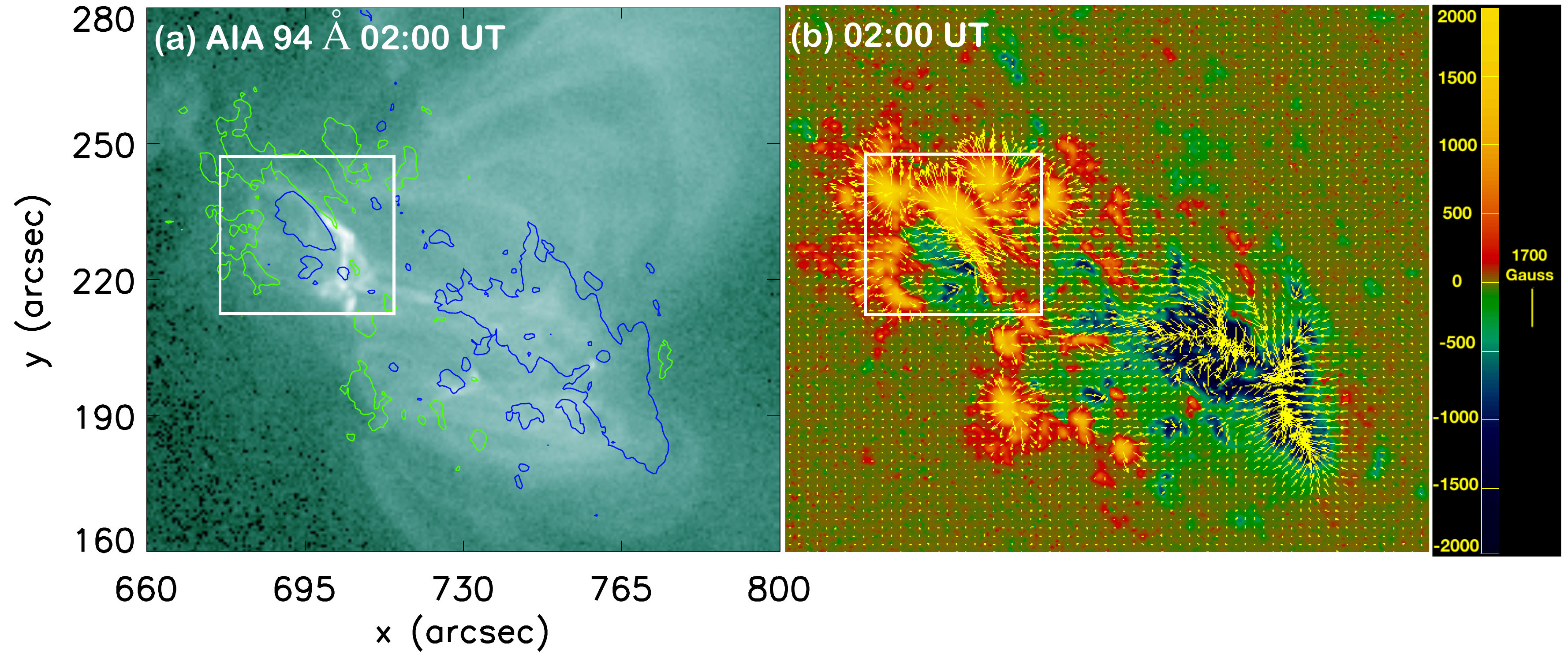}
\caption{Panel (a): Jet base appearing as  brightening with an  arch-shape in between the positive (green contours) and negative (blue contours) in AIA 94 \AA. Panel (b): HMI Vector magnetic field map with the same FOV as of the (a). The white square is the FOV for the Fig. \ref{vector} (a-b). The colour bar indicates the vertical magnetic field strength and the arrow on the right the strength of the horizontal magnetic field in the  map in (b).}
\label{full}
\end{figure*}

Figure \ref{full} presents in the right panel the  magnetic field  vector map  computed  with the  UNNOFIT inversion code 
at 02:00 UT  and the corresponding full AR (left panel)  as  a contextual image meant to  show  the brightening at the base of the jet in AIA 94 \AA.  
 The  vector magnetic field maps represent the  full magnetic field vectors with their three components in the solar local reference frame, 
  generally referred to as the heliospheric reference frame. The vertical component in this reference frame is represented via a colour table. The two horizontal components are associated to form an horizontal vector, which is represented by arrows. However, the pixel dimension is viewed along the line of sight in order to be able to co-align the FOV  with the AIA images.

\begin{figure*}[ht!]
\centering
\includegraphics[width=1.00
\textwidth]{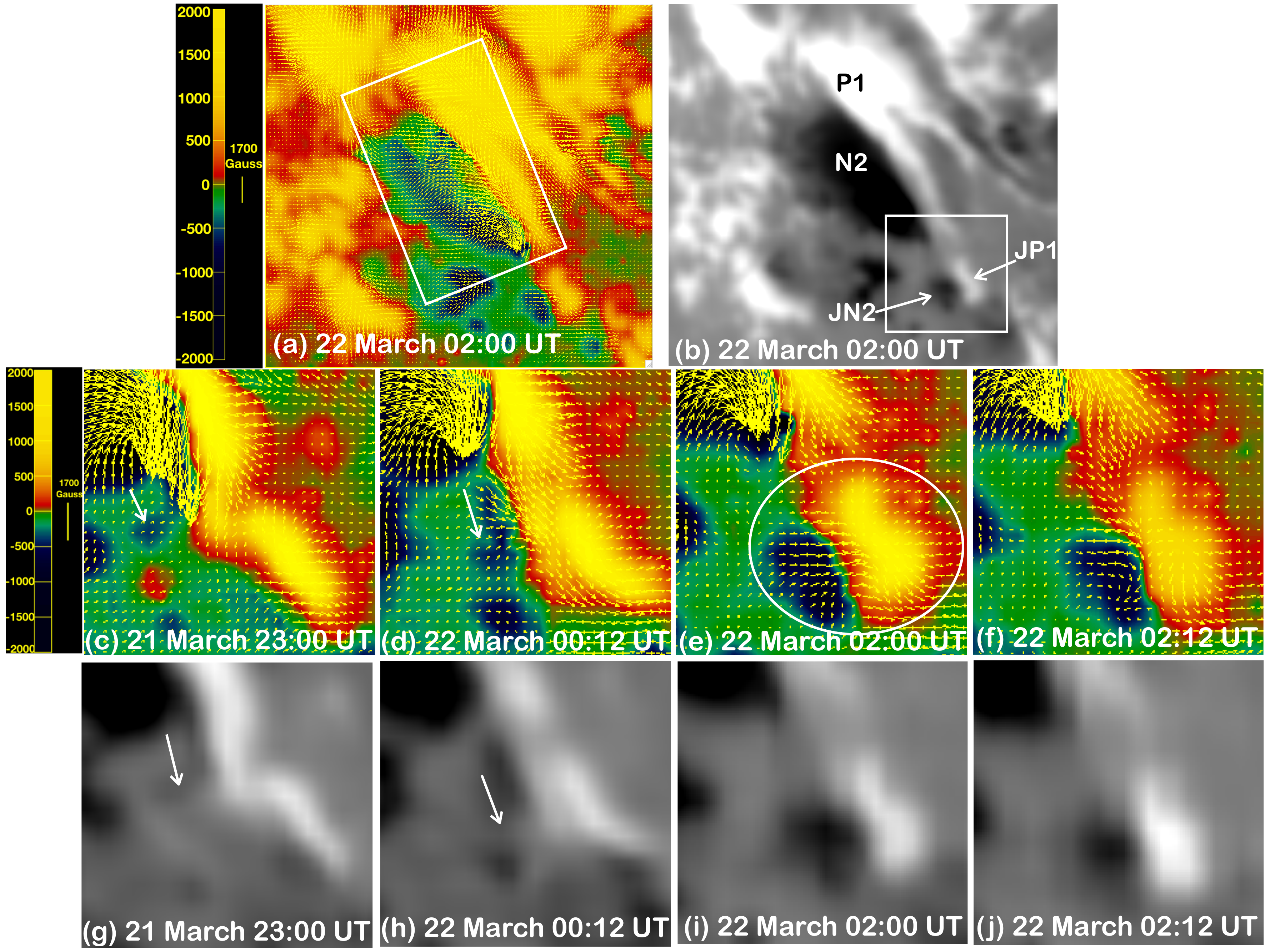}
\caption{Panel (a): Vector magnetic field configuration of a  part of AR NOAA 12736 where the jet is initiated. Panel (b): LOS magnetic configuration at the AR including the two bipoles P1-N2, JP1-JP2. The FOV of panels (a) and (b) is presented as the red square in Fig. \ref{HMI} (e). Panels (c--f) represent the zoomed view of vector magnetic field configuration and panels (g--j) show the zoom view of LOS magnetic field configuration at the small bipole location, where the magnetic flux cancellation occurs. The detached negative (black) polarity is moving towards the south and is shown with white arrows. The FOV of these panels is shown as the white square in panel (b). On the left the colour bar indicates the vertical magnetic field strength and the arrow shows the strength of the horizontal magnetic field in the map (panel (a) and panels (c-f)).}
\label{vector}
\end{figure*}

 A box indicates  the small FOV  encircling the region which contains the brightening at the jet base  corresponding to  the  QSL at the reconnection site. 
We carried out a zoom analysis to probe the nature of magnetic field vectors in this jet region (Fig. \ref{vector} a). 
The length of the arrows represent the strength of the  horizontal magnetic field.  We recognise  the   polarities identified in the longitudinal magnetic field map (Fig. \ref{vector} b).

\subsubsection{Flux rope (FR) vector pattern}

In the long region between P1 and N2,  we  make note of  a  characteristic  pattern of the magnetic field  vectors that  suggests the presence of  
a twisted FR with  vectors converging together in the PIL in the middle part   (between P1 and N2)  and vectors
  turning  at both ends, in the top and bottom parts of the FR, resembling the hooks of a FR (Fig. \ref{vector} (a)).  In the vicinity of the FR,  there is 
  an interface that separates the regions of turning and returning of the vectors,
  which represent the boundary between  the FR  and the arcades over the FR and  its  surrounding area.

This pattern is  relatively stable according with the 22 maps computed around the jet time. On   21 March  at 20:00 UT, the  FR was already created and  was continuously observed until     22 March at 05:00 UT.  At first glance, the FR does  not seem to participate to the  formation of the jet. A very detailed study shows that, in fact, this was a very particular case involving a transfer of  twist from the FR to the jet  during  the FR  extension towards the south  before the   reconnection. This is further  explained in the next section. 

\subsubsection{Formation of  the small bipole}
 
The relationship of the FR and the jet  is detected in the HMI movie of the longitudinal magnetic field where the formation of the small bipole (where the jet was initiated) is observed.
The longitudinal HMI movie (animation attached as MOV1) shows a stress  created by the sliding of  the two opposite polarities (P1 and N2). These two polarities come from the two opposite  magnetic emerging regions (EMF1 and EMF2). On  21 March  at 23:00 UT,  a few hours before the jet, a negative polarity  part of N2  detaches 
and moves towards the south,
sliding along the  positive polarity P1 to form the small bipole JN2-JP1  (Fig. \ref{vector} g-j). The  new bipole  is formed 
with the  small positive (JP1) and  the negative (JN2) polarity
encircled in Fig. \ref{vector} (e). This small bipole  is formed by  collision  of two opposite sign polarities  coming from two  different magnetic systems and not by  direct magnetic flux emergence.

\subsubsection{Bald Patch (BP) magnetic configuration}

Looking at the direction of the magnetic field
vectors  between JN2 and JP1,  we find that they are oriented  from the negative polarity to  the positive polarity, which is evidence that it is a BP region, more generally, that it is a region with magnetic field lines that exhibit  a dip
grazing  the surface at the PIL
(Fig. \ref{vector} e). We note that the BP is observed only at this precise time (02:00 UT) -- and not before  and not after (Fig. \ref{vector}  c, d, f).

\subsubsection{Transfer of twist}

We arrive at the conclusion that with the extension of the FR {towards  the south}, it is possible that the arcades of FR interact with the overlying magnetic field.
Some part of the twist of  the FR could be  transferred to the jet, however, there is still a remnant component in the small bipole, as we see in Fig. \ref{vector} f.  The rotation of the structure at the base could  explain this transfer of twist.
To make certain of the existence of  the FR, we compare this finding with  MHD  simulations.

\subsection{MHD Model}
\label{OHM}
    \subsubsection{Description of the MHD simulation}
    We used  the MHD simulations  of  \citet{Zuccarello2015},  where the physical conditions are used to create a FR in an AR. Starting from an asymmetric, bipolar AR, as in \citet{Aulanier2010},  they investigated different  classes of photospheric motions that are capable of forming  a FR.  Here, we consider the results of the simulations with regard to  converging  motions  towards the PIL of the AR with  magnetic flux cancellation. Progressively twisted magnetic field lines  were  globally  wrapping  around an axis and, eventually, formed a FR. The dynamics of the FR is modelled by using a version of the {\it Observationally driven High-order scheme Magnetohydrodynamic} (OHM) code \citep{Aulanier2005,Aulanier2010}. It is a $\beta$ = 0 simulation, so the plasma conditions are not studied. The MHD simulations have  already been validated by testing different flare activities, such as sigmoid currents  of FR \citep{Aulanier2010}, electric current density increase in flare ribbons \citep{Janvier2014}, and electric current density decrease at CME footpoints
\citep{Barczynski2020}. In this study, we want to test if the footprints of the FR in the HMI magnetic vector (vec B) maps have a similar pattern as  
    the footprints of the  theoretical FR in these MHD simulations.
    
\subsubsection{Comparison between MHD  models and  observations:
    FR pattern of vec B}
    
The comparison between our observations (panels a-b) and MHD simulations (panels c-d) is presented in Figure \ref{model}. We rotated our observation in panel (a-b) by 30$^{\circ}$ in the clockwise direction for an improved comparison with the MHD simulations. It is very clear that a  sheared magnetic field is generated along the PIL and the vectors are strongly inclined along with PIL. We have also evidence for swirling of the magnetic field in the top and bottom part of the FR.

The sheared vec B that converges towards the PIL   is a characteristic motion to create a BP.  This pattern can also be seen in \citet{Barczynski2019}. It is due to the summed effects of: (i) the shear that creates a BP with vec B in the negative  polarity pointing towards the positive polarity; and (ii) the asymmetry of the photospheric flux concentration with a stronger positive polarity (in  the model and observations) which is due to the  magnetic pressure pushing all the fields towards the (weaker) negative  polarity. Hence it is  leading to some sheared vector within the positive  polarity to point  towards the negative polarity. These two effects lead to the convergence.

The swirling motions visible at both ends of FR  are well-represented by vec B, which display  similar angles and  similar spatial gradients  at the edges of the swirlings, which 
separate the swirling vec B from the surrounding magnetic field that has more potential, that is, exhibiting more radial from the center of the magnetic polarity. For example, this separation is visible at the top right in Fig. \ref{model} (a and c) in the positive polarity, where radial vectors are close to turning vectors to the left, and at the  bottom  of the negative polarity, there is a similar separation between radial vectors and vectors turning to the left. This kind of separation is  reminiscent of a QSL, just as in the MHD models (\citep{Janvier2013, Aulanier2019}.  Moreover, 
those swirls correspond  exactly to the footpoints of sigmoidal field lines, even though they are not visible in the extreme ultraviolet (EUV).
Finally, the similarity  of all these characteristics structures 
(e.g. BP, QSL, sigmoidal field line) between the MHD models and our observations leads us to infer the existence of a FR in the immediate vicinity of the jet.

\subsubsection{Comparison between the MHD model and observations: electric current  density (J$_z$) pattern}
 In addition to the pattern of the photospheric horizontal fields, a relatively good match is also found for the vertical current densities.
Both the HMI observation and the MHD simulation display a dominance of the J$_z$ and B$_z$ of the same signs in each polarity of the bipole, with an elongated double-peaked J$_z$ pattern all along the PIL, as well as more extended patches at the ends of the sheared PIL. 

In the model, those extended patches correspond to the footpoints of the FR field lines (Fig. \ref{model} d). One difference, however, between the observation and the model is that with HMI, the extended patches in the negative polarity is more clearly visible than in the positive polarity (see Fig. \ref{model} b). We argue that this difference is minor since it may be due to that fact that in the positive polarity, the swirling patterns of the vector fields are located in relatively weaker vertical fields than in the negative one (i.e. 500G in the former vs. 1500G in the later, see Fig. \ref{model} a). With the same twist in both polarities, the weaker fields result in weaker current densities in the positive polarity. Another difference is that in the MHD simulation, some strong QSL-related current sheets surround the FR footpoint related extended patches (see \citet{Janvier2013, Aulanier2019}. These are not visible with HMI and we argue that this is due to the limitations of the HMI data, from which current sheets can only be extracted during flares and with some processing of the data, as in \citet{Janvier2014}, \citet{Barczynski2020}.
\subsubsection{Comparison between the MHD model and observations: magnitude of J$_{z}$}

\begin{figure*}[ht!]
\centering
\includegraphics[width=1.0
\textwidth]{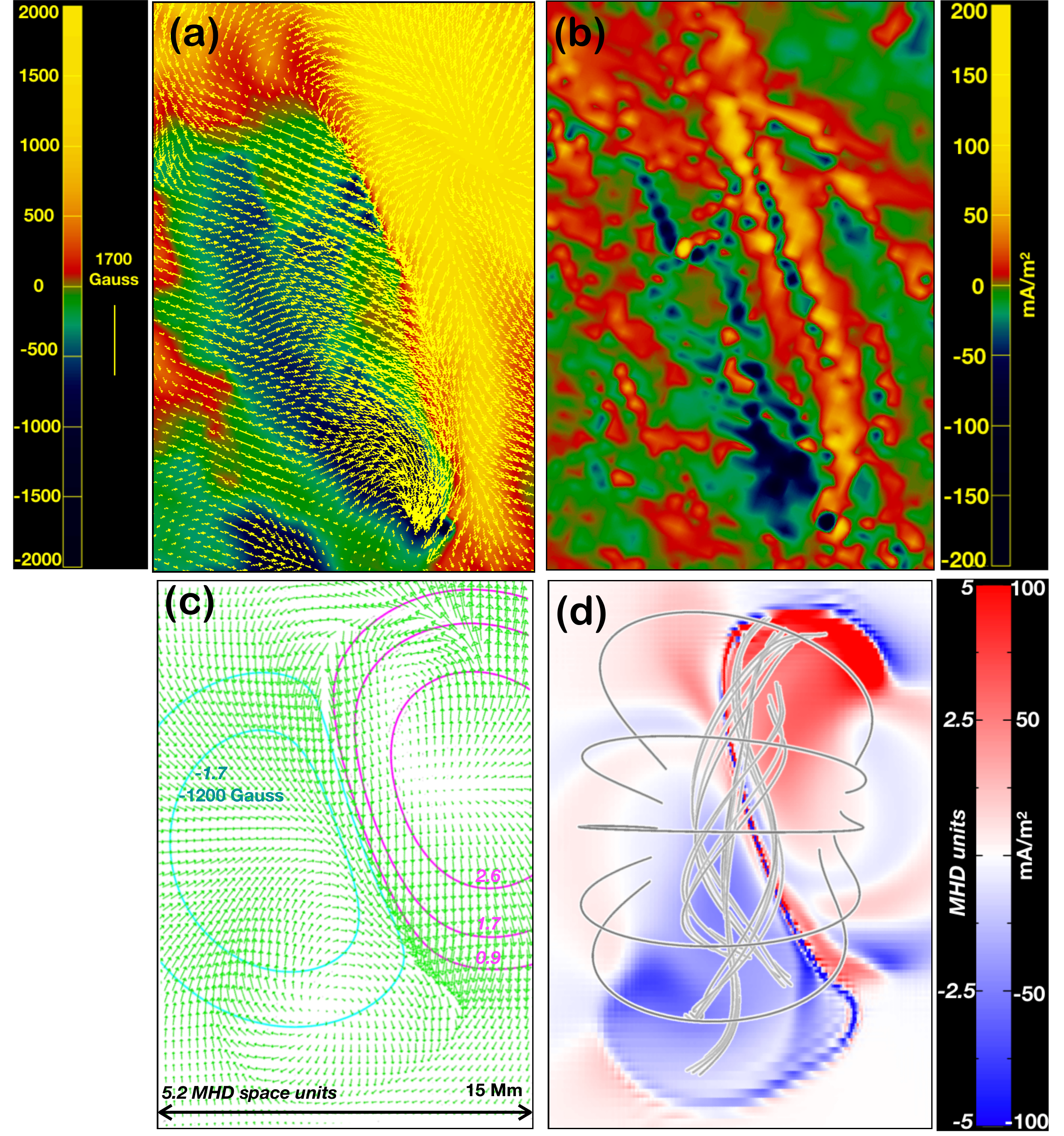}
\caption{Flux rope (FR) evidenced  in the HMI  observations (a,b) and comparison with the images from MHD simulations (c,d). Panel (a): Vector magnetic field map computed with the UNNOFIT code and the yellow (dark) blue areas show the positive (negative) magnetic field polarity. Panel (b): Current density map computed with UNNOFIT code. The FOV for panels (a-b) is presented in Fig. \ref{vector} (a) and to compare with MHD simulation we rotated the observations 30$^{\circ}$ clockwise.  Panel (c): From MHD simulations the  iso-contours of vertical magnetic field with vectors. The pattern of the green vectors is same as with the yellow vectors in observations in panel (a).
Panel (d): The magnetic field lines are plotted with the grey colour and the red and blue contours  are electric currents. So the FR has a very strong electric currents, with the current flowing from red to blue.
The vector pattern of observations and  model looks the same, as they are strongly nearly parallel to the  PIL and converging together in the bottom part. The convergence is due to the asymmetry of the magnetic configuration. The colour bar (top left)  indicates the vertical magnetic field strength in Gauss for panel (a), the colour bar (top right) the strength of electric current in mA m$^{-2}$ for panel (b), the colour bars (bottom right) represent the strength of the electric current in MHD units (left  colour bar), and in physical units (right colour bar) for panel (d).}
\label{model}
\end{figure*}

Comparing the magnitudes of  current densities between  models and  observations requires us to scale the model to physics (solar) units. The reason behind this is that the model was calculated with dimensionless units, resulting in maximum current densities on the order of five units. Such a scaling has already been done for the estimation of flare energies \citep{Aulanier2012,Aulanier2013}. Here, they need to be adjusted to this specific observed bipole. Still, we should bear in mind that this can only be done approximately given the differences in shape between the observed and modelled flux concentrations. Thus, using HMI as a reference (Fig. \ref{model} a), we attributed a magnetic field amplitude of $\pm 1200 G$ to the B$_z$ isocontour $B_z = \pm1.7$ and a bipole size of 15 Mm to the width of 5.2 space units as displayed in Fig. \ref{model} c. Then we scaled the OHM model using a magnetic unit $B_0 = 7.1 \times 10^{-2}$ T and a spatial unit $L_0 = 2.9 \times 10^{6}$ m. We reset the magnetic permeability from unity in the simulation to its real value $\mu$ = 4 $\pi \times 10^{-7}$ SI units. As a result, the  dimensionless current-densities that we model here have to be multiplied by $B_0/(\mu L_0)$ to be expressed in A/m$^{2}$. With these settings, the currents reached up to 100 mA/m$^2$ at the FR footpoints. This value is only half of what is measured with HMI, so the modelled currents are in qualitative agreement with the observed ones. 
The difference in magnitude may be attributed  to the existence of a stronger twist in the observed bipole than  the twist in the model.
Yet it is arguably more likely due the different aspect ratios of the observed and modelled bipoles, the latter being less elongated than the former (compare Fig. \ref{model} a and Fig. \ref{model} c).

\section{Discussion and conclusions} \label{dis}
\subsection{Summary of observations and methodology}
  
The present study concerns multi wavelength observations of   a jet and mini-flare occurring in  the active region (AR) 12736 around 02:04 UT.
We adopted a different methodology than the methodology of \citet{Yang2020}, where the same flare activity was analysed, which resulted in a difference of interpretation.
We looked at the detail history of the polarities and vec B using HMI data instead of the Hinode data, covering the AR, half an hour before the jet. 
The activity was recurrent and evolved very fast with a chain of similar  phases. 
Mini-flares in this AR were  observed frequently changing the connectivity at  this  interface region from a bald patch (BP) region to a current sheet region and vice-versa. The main bipole (P1-N2) is the result of   collision between two emerging fluxes.  The negative polarity is sliding, extending towards  the south, creating a small bipole (JP1-JN2)  and it is cancelled out with the positive polarity.  The history of the AR tells us that JN2 was detached from N2, generating a strong shear along JP1.

Unlike \citet{Yang2020}, we did not make a NLFFF  magnetic extrapolation and, instead, we preferred to  use 
   the horizontal vector magnetic field  (vec B) observations directly to relate the small bipole  and the BP to the origin of the jet (both positioned
at the same place). We compared the observed  magnetic field vec B  pattern  and the values of the electric current density J$_{z}$  with
 synthetic J$_{z}$ and vec B data from the  MHD model of FR \citep{Aulanier2010,Zuccarello2015}  to infer the
location of  FR.  The FR is identified between P1 and N2 in  the HMI vec B maps. It is definitively not the FRs computed  by  the magnetic extrapolation of \citet{Yang2020} neither their first FR (FR1) between N2 and P2 (which, in fact, corresponds to an arch filament), nor their second FR (FR2), with  NS field lines corresponding to a small filament not visible  by absorption in any AIA filters before the jet. The FR1 has
one end in N2 like our FR but the second end is in P2. In our case, the second end (foot)  of the FR 
is in P1,  where we have the signature of a hook in the vec B maps (HMI as well as in Hinode). No similar pattern exists in P2 in the vec B maps.
Our detailed observation analyses suggest that the jet reconnection  occurred in  a  BP current sheet and in the, rapidly formed  above null (`X') point, current sheet,
 driven by the  moving polarity (JN2) that carried twist from the remote FR  and injected it into the jet. The initial FR remains stable during the reconnection process.

\subsection{Scenario of transfer of twist (cartoon)}

\begin{figure*}[ht!]
\centering
\includegraphics[width=0.80
\textwidth]{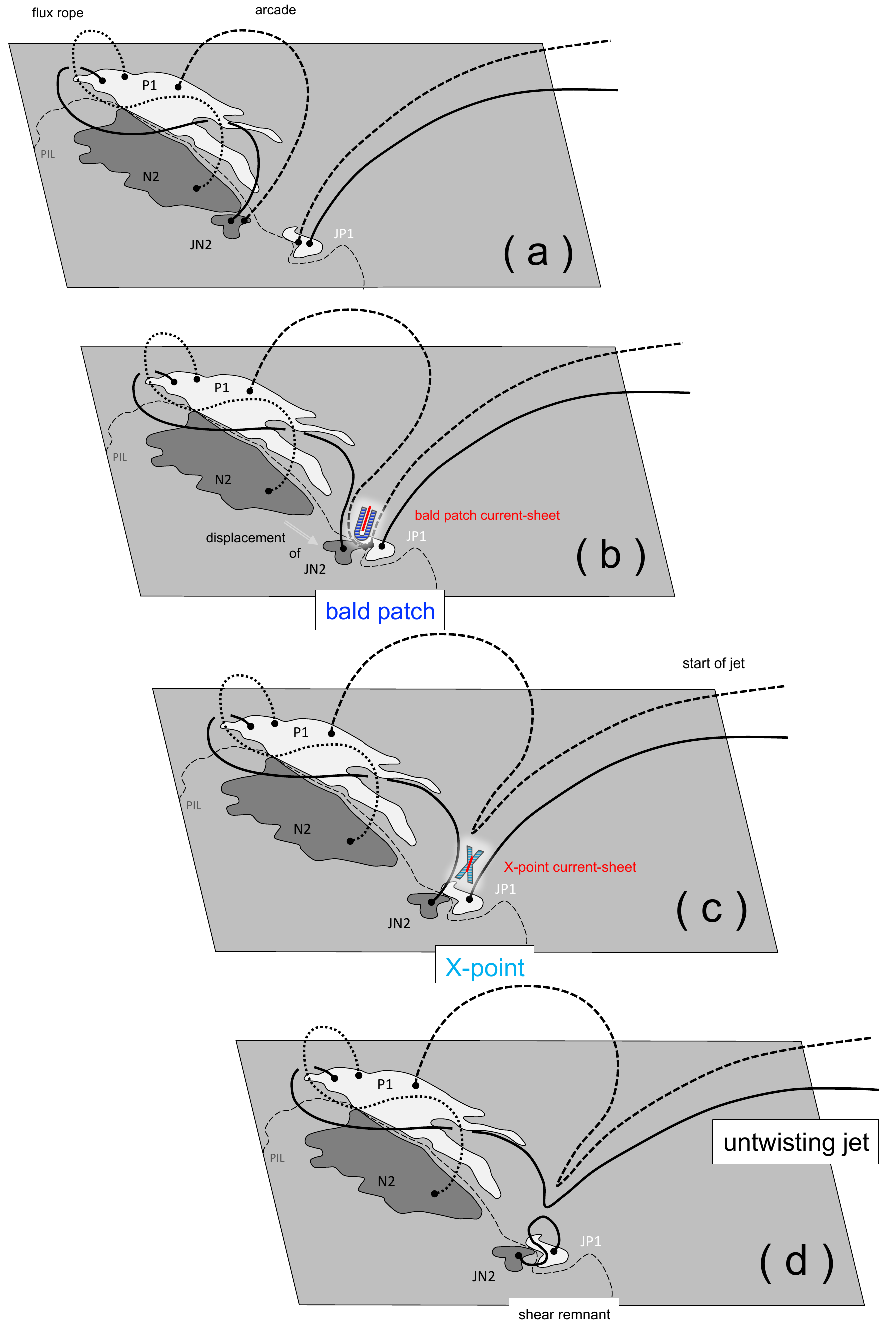}
\caption{Sketch  of the formation of the jet and  transfer of the twist from the FR to the jet during reconnection.
Panel (a) shows the magnetic configuration before  the reconnection, panel (b)  shows the formation of the BP current sheet, panel (c) shows the X-point current sheet, panel (d) shows the untwisting jet after the reconnection and the remnant twist in the bipole JP1 and JN2.
The dashed lines represent the non-eruptive FR, the arcade between P1 and N2 (dashed line in a)  reconnects at the BP with the dashed line on the right in  (b) creating a long dashed field lines from P1 to the extreme right. Below the arcade, the twisted line P1-N2 of  FR  (solid line)  is elongated in (b) and touches the BP region in (c), it  reconnects with  an  open magnetic field line (solid line  on the right in c)  creating the untwisting jet in (d).}
\label{cartoon}
\end{figure*}

We proposed a cartoon
where the FR between P1 and N2 is represented by the solid  twisted line  (Fig. \ref{cartoon}, a). 
It is extended to the  south, creating the bipole JN2-JP1. The BP  current sheet is generated  between the overlying arcade of FR and the magnetic field line of the west  emerging flux P1-N1 (panel b).  At this time, a first reconnection occurs at a localised point that is very deep in the atmosphere. 
The Mg II  profiles resemble  those found in IRIS bombs (IB)  with extended wings \citep{Peter2014}, which are proposed to have been formed during   BP current sheet reconnection \citep{Zhao2017}. 
Such chromospheric wide profiles   have been  modelled in MHD simulations   \citep{Hansteen2019}.
It has also been shown that a BP could be  transformed immediately at a  null point.
We propose in panel (c) that  the reconnection occurs in the null point (`X'-point) that is formed dynamically  along a  current sheet  or  a flat spine-surface  above a dome that is not depicted in the cartoon  panel (c). Cool material trapped in the BP during its formation is  expelled with a large blueshift,  
as revealed in IRIS Mg II line profiles with extended blue  wings. The spectra shows an evident tilt, which indicates the presence of helical motions.   The reconnection  site  is heated at all the temperatures and the  hot jet is expelled towards the west  side in twisted field lines (panel d). The cool material follows different paths than the hot and acts as a wall in front of the hot jet. It resembles the  surges that accompany  jets in the MHD simulations of \citet{Moreno2008}, \citet{Nobrega2016}, and \citet{Nobrega2018}.
The HMI vec B map  shows some remnant twist in the bipole JP1 and JN2  after the ejection (panel d), which offers evidence that the twist has gone mainly into the ejected jet.

\subsection{Scenario of the breakout}

The scenario proposed by \citet{Yang2020} is based on their own  NLFFF magnetic  extrapolation, which  suggests that the  small FR2 erupted in the breakout scenario with
a reconnection
at a null point \citep{Antiochos1999}.
We agree globally to the NLFFF extrapolation with the existence of a null point, fan, and spine. The QSLs are well defined and could correspond to the base of the dome as we describe as QSLs are robust structures but their precise location is difficult to determine accurately and to thus serve as an argument for localising filaments \citep{Dalmasse2015,Joshi2019}.

The role of FR2 is important in their case,  with its strong pre-existing shear.  They argue that it is an instability breakout-related jet 
leading to a full evacuation of their FR2 \citep{Sterling2016,Wyper2017}. However, the identification of their FRs in the observations is difficult in AIA images, where there are mixed bright and dark paths in this zone (Fig. \ref{AIA_IRIS}), and even in the images of NVST.  In addition, the ejection of the cool material is blue-shifted, so it runs in the opposite direction   to the hot jet. It is difficult to trust that  the eruption of this cool material is the driver of  the jet.

 Moreover, in their  NLFFF magnetic extrapolation  the null point is located quite
far
from   the tiny bipole JN2-JP1 (see  Fig. 3 in their paper), which is, hence, far from the observed origin of the jet. 
This  location of the X point is may be due to the fact that the authors worked with an earlier  vect B map from nearly half an hour before the reconnection. It could be the location of the former X point of the previous 
mini-flare, which is when JN2 was not yet colliding with JP1. The BP was not yet formed at this early stage. We show that it is important to take into account  the history of this AR, which evolved quite fast. This difference in the localisation of the null point  could be due to  the nonalignment of HMI, AIA,  and IRIS  images in the paper of \citet{Yang2020}, where the authors do not   note   that the  nominal coordinates of 
these  instruments have 4 arcsec in x, 3 arcsec of y difference and, instead, they mention a projection effect.

\subsection{Conclusions}

In this paper, we present the observations of a twisted jet, a surge, and a mini-flare which occurred  in the active region (AR) 12736 on 22 March 2019 at 02:05UT. The event was observed in multi-wavelengths with AIA and IRIS instruments and detailed in the magnetic  field vector maps obtained by HMI and  computed with the UNNOFIT code. The MHD simulations were used to validate the vec B observations \citep{Aulanier2010,Zuccarello2015}.
We present our main results in the following :
\begin{enumerate}
\item The AR consisted of the collapse of two  emerging magnetic flux (EMF) regions, each of them overlaid by an  arch filament system (AFS). The jet and surge  reconnection site is along the PIL between these two AFS. The AFS over the east side  evolved rapidly due to photospheric surface motions. Prior to the reconnection, the AFS exhibit a dome shape. After the reconnection, long AFS overlying both EMFs are observed. This  is confirmed in  the NVST H$\alpha$ images.

\item 
A large flux rope (FR) in the vicinity of the jet region is detected. The patterns of transverse fields and vertical current densities, as observed by HMI and appearing without being constrained  a priori in an MHD simulation of non-eruptive 
FR  formation with flux-cancellation of sheared loops, show a  good accordance.
The location of the FR  is  fully supported by HMI vec B  and electric currents J$_z$ maps.

\item 
The magnetic topology of the AR demonstrates  a bald patch (BP)  region  due to the particular formation of the bipole by collision of opposite polarities, which is  dynamically  transformed to an `X'-point  current sheet.

\item 
The fast extension of the FR towards the site of reconnection  due to photospheric surface motions offers the possibility for the FR arcades to reconnect with magnetic pre-existing field lines at the `X'-point current sheet without the eruption of the  FR. The extension of the FR may transmit twist to the jet.

\item 
The IRIS spectra at the reconnection site display a tilt and gradient in the spectra along the jet base, indicating the formation of a rotating structure during the reconnection.

\end{enumerate}

The clues 
leading to our interpretation consists of the identification of a non-eruptive FR, from which some twist is carried away and eventually reconnected into the jet at the 
`X'-point current-sheet.
The transport of twist away from the FR towards a BP is supported by the HMI observations of a moving negative flux-concentration whose transverse fields point towards a positive one. The twist is transported at a long distance of the FR which remains non-eruptive. The tilt observed in the IRIS spectra in the four positions of the slit, which, by chance, are exactly at the site reconnection,  confirms the transfer of twist at the jet base.

Our magnetic analysis benefit from the treatment of the HMI vec B by the UNNOFIT code, which uses a filling factor that takes into account the non-resolved structures. In each pixel, there is an equilibrium between magnetised regions and non-magnetised regions, which implies a better determination of the magnetic field inclination \citep{Bommier2016}. This is an important aspect  for  regions with a weak magnetic field. This is the case in the small bipole, where our jet  reconnection takes place  and where we have detected the BP.
The second aspect that is important in this study is the chance to have the IRIS spectra at the right place, namely, at the reconnection site. The IRIS spectra directly shows  the transfer of twist between two stable systems at the reconnection point by unveiling a helical structure.

 Observations of bright jets simultaneously followed with cool and dense material or surges are an interesting area of study to probe by  spectroscopic analysis using the  IRIS instrument.   The IRIS spectral profiles  are  very useful tools for explaining dynamics, such as the  rotation in jets,  as we show here. They may also open up a new field of study to determine  the physical plasma conditions  of  the atmosphere at the reconnection site that is directly related to mini-flares. 
 
\begin{acknowledgements}
We are grateful to the anonymous referee for their valuable comments.
We thank the SDO/AIA, SDO/HMI, and IRIS science teams for granting free access to the data. 
The H$\alpha$ data used in this paper were obtained with the New Vacuum Solar Telescope in Fuxian Solar Observatory of Yunnan Astronomical Observatory, CAS. We thank Xiang Yong Gyuan  for providing us the level 1 data.
 This work is carried out at Observatorie de Paris, Meudon, France under the Raman Charpak fellowship by CEFIPRA. RJ thanks to CEFIPRA for a Raman Charpak fellowship and to the Department of Science and Technology New Delhi, India for the INSPIRE fellowship. 
This work was granted access to the HPC resources of MesoPSL financed by the Region Ile de France
and the project Equip@Meso (reference ANR-10-EQPX-29-01) of the \textit{Investissements d'Avenir} program supervised by the \textit{Agence Nationale pour la Recherche}. The work of RC is supported from the 
Bulgarian Science Fund under Indo-Bulgarian bilateral project. All authors thank to the providers of open-source software for online
calls and meetings, which were essential for the completion
of this work during the outbreak of the COVID-19
pandemic.
\end{acknowledgements}

\bibliography{references-4}

\begin{thebibliography}{80}
\expandafter\ifx\csname natexlab\endcsname\relax\def\natexlab#1{#1}\fi

\bibitem[{{Alissandrakis} {et~al.}(2018){Alissandrakis}, {Vial}, {Koukras},
  {Buchlin}, \& {Chane-Yook}}]{Alissandrakis2018}
{Alissandrakis}, C.~E., {Vial}, J.-C., {Koukras}, A., {Buchlin}, E., \&
  {Chane-Yook}, M. 2018, \solphys, 293, 20

\bibitem[{{Antiochos} \& {DeVore}(1999)}]{Antiochos1999}
{Antiochos}, S.~K. \& {DeVore}, C.~R. 1999, Washington DC American Geophysical
  Union Geophysical Monograph Series, 199, 113

\bibitem[{{Anzer} \& {Heinzel}(2005)}]{Anzer2005}
{Anzer}, U. \& {Heinzel}, P. 2005, \apj, 622, 714

\bibitem[{{Archontis} {et~al.}(2004){Archontis}, {Moreno-Insertis},
  {Galsgaard}, {Hood}, \& {O'Shea}}]{Archontis2004}
{Archontis}, V., {Moreno-Insertis}, F., {Galsgaard}, K., {Hood}, A., \&
  {O'Shea}, E. 2004, \aap, 426, 1047

\bibitem[{{Archontis} {et~al.}(2005){Archontis}, {Moreno-Insertis},
  {Galsgaard}, \& {Hood}}]{Archontis2005}
{Archontis}, V., {Moreno-Insertis}, F., {Galsgaard}, K., \& {Hood}, A.~W. 2005,
  \apj, 635, 1299

\bibitem[{{Asai} {et~al.}(2001){Asai}, {Ishii}, \& {Kurokawa}}]{Asai2001}
{Asai}, A., {Ishii}, T.~T., \& {Kurokawa}, H. 2001, \apjl, 555, L65

\bibitem[{{Aulanier} {et~al.}(2013){Aulanier}, {D{\'e}moulin}, {Schrijver},
  {Janvier}, {Pariat}, \& {Schmieder}}]{Aulanier2013}
{Aulanier}, G., {D{\'e}moulin}, P., {Schrijver}, C.~J., {et~al.} 2013, \aap,
  549, A66

\bibitem[{{Aulanier} \& {Dud{\'\i}k}(2019)}]{Aulanier2019}
{Aulanier}, G. \& {Dud{\'\i}k}, J. 2019, \aap, 621, A72

\bibitem[{{Aulanier} {et~al.}(2012){Aulanier}, {Janvier}, \&
  {Schmieder}}]{Aulanier2012}
{Aulanier}, G., {Janvier}, M., \& {Schmieder}, B. 2012, \aap, 543, A110

\bibitem[{{Aulanier} {et~al.}(2005){Aulanier}, {Pariat}, \&
  {D{\'e}moulin}}]{Aulanier2005}
{Aulanier}, G., {Pariat}, E., \& {D{\'e}moulin}, P. 2005, \aap, 444, 961

\bibitem[{{Aulanier} {et~al.}(2010){Aulanier}, {T{\"o}r{\"o}k}, {D{\'e}moulin},
  \& {DeLuca}}]{Aulanier2010}
{Aulanier}, G., {T{\"o}r{\"o}k}, T., {D{\'e}moulin}, P., \& {DeLuca}, E.~E.
  2010, \apj, 708, 314

\bibitem[{{Barczynski} {et~al.}(2020){Barczynski}, {Aulanier}, {Janvier},
  {Schmieder}, \& {Masson}}]{Barczynski2020}
{Barczynski}, K., {Aulanier}, G., {Janvier}, M., {Schmieder}, B., \& {Masson},
  S. 2020, \apj, 895, 18

\bibitem[{{Barczynski} {et~al.}(2019){Barczynski}, {Aulanier}, {Masson}, \&
  {Wheatland}}]{Barczynski2019}
{Barczynski}, K., {Aulanier}, G., {Masson}, S., \& {Wheatland}, M.~S. 2019,
  \apj, 877, 67

\bibitem[{{Bernasconi} {et~al.}(2002){Bernasconi}, {Rust}, {Georgoulis}, \&
  {Labonte}}]{Bernasconi2002}
{Bernasconi}, P.~N., {Rust}, D.~M., {Georgoulis}, M.~K., \& {Labonte}, B.~J.
  2002, \solphys, 209, 119

\bibitem[{{Bommier}(2016)}]{Bommier2016}
{Bommier}, V. 2016, Journal of Geophysical Research (Space Physics), 121, 5025

\bibitem[{{Bommier} {et~al.}(2007){Bommier}, {Landi Degl'Innocenti},
  {Landolfi}, \& {Molodij}}]{Bommier2007}
{Bommier}, V., {Landi Degl'Innocenti}, E., {Landolfi}, M., \& {Molodij}, G.
  2007, \aap, 464, 323

\bibitem[{{Canfield} {et~al.}(1996){Canfield}, {Reardon}, {Leka}, {Shibata},
  {Yokoyama}, \& {Shimojo}}]{Canfield1996}
{Canfield}, R.~C., {Reardon}, K.~P., {Leka}, K.~D., {et~al.} 1996, \apj, 464,
  1016

\bibitem[{{Chae} {et~al.}(1999){Chae}, {Qiu}, {Wang}, \& {Goode}}]{Chae1999}
{Chae}, J., {Qiu}, J., {Wang}, H., \& {Goode}, P.~R. 1999, \apjl, 513, L75

\bibitem[{{Chandra} {et~al.}(2017){Chandra}, {Mandrini}, {Schmieder}, {Joshi},
  {Cristiani}, {Cremades}, {Pariat}, {Nuevo}, {Srivastava}, \&
  {Uddin}}]{Chandra2017}
{Chandra}, R., {Mandrini}, C.~H., {Schmieder}, B., {et~al.} 2017, \aap, 598,
  A41

\bibitem[{{Chen} {et~al.}(2012){Chen}, {Zhang}, \& {Ma}}]{Chen2012}
{Chen}, H.-D., {Zhang}, J., \& {Ma}, S.-L. 2012, Research in Astronomy and
  Astrophysics, 12, 573

\bibitem[{{Curdt} {et~al.}(2012){Curdt}, {Tian}, \& {Kamio}}]{Curdt2012}
{Curdt}, W., {Tian}, H., \& {Kamio}, S. 2012, \solphys, 280, 417

\bibitem[{{Dalmasse} {et~al.}(2015){Dalmasse}, {Chandra}, {Schmieder}, \&
  {Aulanier}}]{Dalmasse2015}
{Dalmasse}, K., {Chandra}, R., {Schmieder}, B., \& {Aulanier}, G. 2015, \aap,
  574, A37

\bibitem[{{De Pontieu} {et~al.}(2014){De Pontieu}, {Title}, {Lemen}, {Kushner},
  {Akin}, {Allard}, {Berger}, {Boerner}, {Cheung}, {Chou}, {Drake}, {Duncan},
  {Freeland}, {Heyman}, {Hoffman}, {Hurlburt}, {Lindgren}, {Mathur}, {Rehse},
  {Sabolish}, {Seguin}, {Schrijver}, {Tarbell}, {W{\"u}lser}, {Wolfson},
  {Yanari}, {Mudge}, {Nguyen-Phuc}, {Timmons}, {van Bezooijen}, {Weingrod},
  {Brookner}, {Butcher}, {Dougherty}, {Eder}, {Knagenhjelm}, {Larsen},
  {Mansir}, {Phan}, {Boyle}, {Cheimets}, {DeLuca}, {Golub}, {Gates}, {Hertz},
  {McKillop}, {Park}, {Perry}, {Podgorski}, {Reeves}, {Saar}, {Testa}, {Tian},
  {Weber}, {Dunn}, {Eccles}, {Jaeggli}, {Kankelborg}, {Mashburn}, {Pust},
  {Springer}, {Carvalho}, {Kleint}, {Marmie}, {Mazmanian}, {Pereira}, {Sawyer},
  {Strong}, {Worden}, {Carlsson}, {Hansteen}, {Leenaarts}, {Wiesmann},
  {Aloise}, {Chu}, {Bush}, {Scherrer}, {Brekke}, {Martinez-Sykora}, {Lites},
  {McIntosh}, {Uitenbroek}, {Okamoto}, {Gummin}, {Auker}, {Jerram}, {Pool}, \&
  {Waltham}}]{Pontieu2014}
{De Pontieu}, B., {Title}, A.~M., {Lemen}, J.~R., {et~al.} 2014, \solphys, 289,
  2733

\bibitem[{{D{\'e}moulin} {et~al.}(1996){D{\'e}moulin}, {Henoux}, {Priest}, \&
  {Mandrini}}]{Demoulin1996}
{D{\'e}moulin}, P., {Henoux}, J.~C., {Priest}, E.~R., \& {Mandrini}, C.~H.
  1996, \aap, 308, 643

\bibitem[{{Georgoulis} {et~al.}(2002){Georgoulis}, {Rust}, {Bernasconi}, \&
  {Schmieder}}]{Georgoulis2002}
{Georgoulis}, M.~K., {Rust}, D.~M., {Bernasconi}, P.~N., \& {Schmieder}, B.
  2002, \apj, 575, 506

\bibitem[{{Grubecka} {et~al.}(2016){Grubecka}, {Schmieder}, {Berlicki},
  {Heinzel}, {Dalmasse}, \& {Mein}}]{Grubecka2016}
{Grubecka}, M., {Schmieder}, B., {Berlicki}, A., {et~al.} 2016, \aap, 593, A32

\bibitem[{{Gu} {et~al.}(1994){Gu}, {Lin}, {Li}, {Xuan}, {Luan}, \&
  {Li}}]{Gu1994}
{Gu}, X.~M., {Lin}, J., {Li}, K.~J., {et~al.} 1994, \aap, 282, 240

\bibitem[{{Hansteen} {et~al.}(2019){Hansteen}, {Ortiz}, {Archontis},
  {Carlsson}, {Pereira}, \& {Bj{\o}rgen}}]{Hansteen2019}
{Hansteen}, V., {Ortiz}, A., {Archontis}, V., {et~al.} 2019, \aap, 626, A33

\bibitem[{{Heyvaerts} {et~al.}(1977){Heyvaerts}, {Priest}, \&
  {Rust}}]{Heyvaerts1977}
{Heyvaerts}, J., {Priest}, E.~R., \& {Rust}, D.~M. 1977, \apj, 216, 123

\bibitem[{{Hong} {et~al.}(2013){Hong}, {Jiang}, {Yang}, {Zheng}, {Bi}, {Li},
  {Yang}, \& {Yang}}]{Hong2013}
{Hong}, J.-C., {Jiang}, Y.-C., {Yang}, J.-Y., {et~al.} 2013, Research in
  Astronomy and Astrophysics, 13, 253

\bibitem[{{Janvier} {et~al.}(2014){Janvier}, {Aulanier}, {Bommier},
  {Schmieder}, {D{\'e}moulin}, \& {Pariat}}]{Janvier2014}
{Janvier}, M., {Aulanier}, G., {Bommier}, V., {et~al.} 2014, \apj, 788, 60

\bibitem[{{Janvier} {et~al.}(2013){Janvier}, {Aulanier}, {Pariat}, \&
  {D{\'e}moulin}}]{Janvier2013}
{Janvier}, M., {Aulanier}, G., {Pariat}, E., \& {D{\'e}moulin}, P. 2013, \aap,
  555, A77

\bibitem[{{Joshi} {et~al.}(2019){Joshi}, {Zhu}, {Schmieder}, {Aulanier},
  {Janvier}, {Joshi}, {Magara}, {Chandra}, \& {Inoue}}]{Joshi2019}
{Joshi}, N.~C., {Zhu}, X., {Schmieder}, B., {et~al.} 2019, \apj, 871, 165

\bibitem[{{Joshi} {et~al.}(2020){Joshi}, {Chandra}, {Schmieder},
  {Moreno-Insertis}, {Aulanier}, {N{\'o}brega-Siverio}, \& {Devi}}]{Joshi2020}
{Joshi}, R., {Chandra}, R., {Schmieder}, B., {et~al.} 2020, \aap, 639, A22

\bibitem[{{Joshi} {et~al.}(2017){Joshi}, {Schmieder}, {Chandra}, {Aulanier},
  {Zuccarello}, \& {Uddin}}]{Joshi2017b}
{Joshi}, R., {Schmieder}, B., {Chandra}, R., {et~al.} 2017, \solphys, 292, 152

\bibitem[{{Leka} {et~al.}(2009){Leka}, {Barnes}, {Crouch}, {Metcalf}, {Gary},
  {Jing}, \& {Liu}}]{Leka2009}
{Leka}, K.~D., {Barnes}, G., {Crouch}, A.~D., {et~al.} 2009, \solphys, 260, 83

\bibitem[{{Lemen} {et~al.}(2012){Lemen}, {Title}, {Akin}, {Boerner}, {Chou},
  {Drake}, {Duncan}, {Edwards}, {Friedlaender}, {Heyman}, {Hurlburt}, {Katz},
  {Kushner}, {Levay}, {Lindgren}, {Mathur}, {McFeaters}, {Mitchell}, {Rehse},
  {Schrijver}, {Springer}, {Stern}, {Tarbell}, {Wuelser}, {Wolfson}, {Yanari},
  {Bookbinder}, {Cheimets}, {Caldwell}, {Deluca}, {Gates}, {Golub}, {Park},
  {Podgorski}, {Bush}, {Scherrer}, {Gummin}, {Smith}, {Auker}, {Jerram},
  {Pool}, {Soufli}, {Windt}, {Beardsley}, {Clapp}, {Lang}, \&
  {Waltham}}]{Lemen2012}
{Lemen}, J.~R., {Title}, A.~M., {Akin}, D.~J., {et~al.} 2012, \solphys, 275, 17

\bibitem[{{Li} {et~al.}(2018){Li}, {Li}, \& {Ning}}]{Li2018}
{Li}, D., {Li}, L., \& {Ning}, Z. 2018, \mnras, 479, 2382

\bibitem[{{Li} {et~al.}(2014){Li}, {Peter}, {Chen}, \& {Zhang}}]{Li2014}
{Li}, L.~P., {Peter}, H., {Chen}, F., \& {Zhang}, J. 2014, \aap, 570, A93

\bibitem[{{Liu} {et~al.}(2014){Liu}, {Xu}, {Gu}, {Wang}, {You}, {Shen}, {Lu},
  {Jin}, {Chen}, {Lou}, {Li}, {Liu}, {Xu}, {Rao}, {Hu}, {Li}, {Fu}, {Wang},
  {Bao}, {Wu}, \& {Zhang}}]{NVST2014}
{Liu}, Z., {Xu}, J., {Gu}, B.-Z., {et~al.} 2014, Research in Astronomy and
  Astrophysics, 14, 705

\bibitem[{{Mandrini} {et~al.}(2002){Mandrini}, {D{\'e}moulin}, {Schmieder},
  {Deng}, \& {Rudawy}}]{Mandrini2002}
{Mandrini}, C.~H., {D{\'e}moulin}, P., {Schmieder}, B., {Deng}, Y.~Y., \&
  {Rudawy}, P. 2002, \aap, 391, 317

\bibitem[{{Moore} {et~al.}(2010){Moore}, {Cirtain}, {Sterling}, \&
  {Falconer}}]{Moore2010}
{Moore}, R.~L., {Cirtain}, J.~W., {Sterling}, A.~C., \& {Falconer}, D.~A. 2010,
  \apj, 720, 757

\bibitem[{{Moreno-Insertis} \& {Galsgaard}(2013)}]{Moreno2013}
{Moreno-Insertis}, F. \& {Galsgaard}, K. 2013, \apj, 771, 20

\bibitem[{{Moreno-Insertis} {et~al.}(2008){Moreno-Insertis}, {Galsgaard}, \&
  {Ugarte-Urra}}]{Moreno2008}
{Moreno-Insertis}, F., {Galsgaard}, K., \& {Ugarte-Urra}, I. 2008, \apjl, 673,
  L211

\bibitem[{{Nistic{\`o}} {et~al.}(2009){Nistic{\`o}}, {Bothmer}, {Patsourakos},
  \& {Zimbardo}}]{Nistico2009}
{Nistic{\`o}}, G., {Bothmer}, V., {Patsourakos}, S., \& {Zimbardo}, G. 2009,
  \solphys, 259, 87

\bibitem[{{N{\'o}brega-Siverio} {et~al.}(2017){N{\'o}brega-Siverio},
  {Mart{\'{\i}}nez-Sykora}, {Moreno-Insertis}, \& {Rouppe van der
  Voort}}]{Nobrega2017}
{N{\'o}brega-Siverio}, D., {Mart{\'{\i}}nez-Sykora}, J., {Moreno-Insertis}, F.,
  \& {Rouppe van der Voort}, L. 2017, \apj, 850, 153

\bibitem[{{N{\'o}brega-Siverio} {et~al.}(2016){N{\'o}brega-Siverio},
  {Moreno-Insertis}, \& {Mart{\'\i}nez-Sykora}}]{Nobrega2016}
{N{\'o}brega-Siverio}, D., {Moreno-Insertis}, F., \& {Mart{\'\i}nez-Sykora}, J.
  2016, \apj, 822, 18

\bibitem[{{N{\'o}brega-Siverio} {et~al.}(2018){N{\'o}brega-Siverio},
  {Moreno-Insertis}, \& {Mart{\'\i}nez-Sykora}}]{Nobrega2018}
{N{\'o}brega-Siverio}, D., {Moreno-Insertis}, F., \& {Mart{\'\i}nez-Sykora}, J.
  2018, \apj, 858, 8

\bibitem[{{Pariat} {et~al.}(2010){Pariat}, {Antiochos}, \&
  {DeVore}}]{Pariat2010}
{Pariat}, E., {Antiochos}, S.~K., \& {DeVore}, C.~R. 2010, \apj, 714, 1762

\bibitem[{{Pariat} {et~al.}(2015){Pariat}, {Dalmasse}, {DeVore}, {Antiochos},
  \& {Karpen}}]{Pariat2015}
{Pariat}, E., {Dalmasse}, K., {DeVore}, C.~R., {Antiochos}, S.~K., \& {Karpen},
  J.~T. 2015, \aap, 573, A130

\bibitem[{{Pariat} {et~al.}(2016){Pariat}, {Dalmasse}, {DeVore}, {Antiochos},
  \& {Karpen}}]{Pariat2016}
{Pariat}, E., {Dalmasse}, K., {DeVore}, C.~R., {Antiochos}, S.~K., \& {Karpen},
  J.~T. 2016, \aap, 596, A36

\bibitem[{{Patsourakos} {et~al.}(2008){Patsourakos}, {Pariat}, {Vourlidas},
  {Antiochos}, \& {Wuelser}}]{Patsourakos2008}
{Patsourakos}, S., {Pariat}, E., {Vourlidas}, A., {Antiochos}, S.~K., \&
  {Wuelser}, J.~P. 2008, \apjl, 680, L73

\bibitem[{{Pesnell} {et~al.}(2012){Pesnell}, {Thompson}, \&
  {Chamberlin}}]{Pesnell2012}
{Pesnell}, W.~D., {Thompson}, B.~J., \& {Chamberlin}, P.~C. 2012, \solphys,
  275, 3

\bibitem[{{Peter} {et~al.}(2014){Peter}, {Tian}, {Curdt}, {Schmit}, {Innes},
  {De Pontieu}, {Lemen}, {Title}, {Boerner}, {Hurlburt}, {Tarbell}, {Wuelser},
  {Mart{\'{\i}}nez-Sykora}, {Kleint}, {Golub}, {McKillop}, {Reeves}, {Saar},
  {Testa}, {Kankelborg}, {Jaeggli}, {Carlsson}, \& {Hansteen}}]{Peter2014}
{Peter}, H., {Tian}, H., {Curdt}, W., {et~al.} 2014, Science, 346, 1255726

\bibitem[{{Priest} {et~al.}(2018){Priest}, {Chitta}, \&
  {Syntelis}}]{Priest2018}
{Priest}, E.~R., {Chitta}, L.~P., \& {Syntelis}, P. 2018, \apjl, 862, L24

\bibitem[{{Raouafi} {et~al.}(2016){Raouafi}, {Patsourakos}, {Pariat}, {Young},
  {Sterling}, {Savcheva}, {Shimojo}, {Moreno-Insertis}, {DeVore}, {Archontis},
  {T{\"o}r{\"o}k}, {Mason}, {Curdt}, {Meyer}, {Dalmasse}, \&
  {Matsui}}]{Raouafi2016}
{Raouafi}, N.~E., {Patsourakos}, S., {Pariat}, E., {et~al.} 2016, \ssr, 201, 1

\bibitem[{{Rompolt}(1975)}]{Rompolt1975}
{Rompolt}, B. 1975, \solphys, 41, 329

\bibitem[{{Ruan} {et~al.}(2019){Ruan}, {Schmieder}, {Masson}, {Mein}, {Mein},
  G., \& {Chen}}]{Ruan2019}
{Ruan}, G., {Schmieder}, B., {Masson}, S., {et~al.} 2019, \apj, 0, 0

\bibitem[{{Schmieder} {et~al.}(1994){Schmieder}, {Golub}, \&
  {Antiochos}}]{Schmieder1994b}
{Schmieder}, B., {Golub}, L., \& {Antiochos}, S.~K. 1994, \apj, 425, 326

\bibitem[{{Schmieder} {et~al.}(2013){Schmieder}, {Guo}, {Moreno-Insertis},
  {Aulanier}, {Yelles Chaouche}, {Nishizuka}, {Harra}, {Thalmann}, {Vargas
  Dominguez}, \& {Liu}}]{Schmieder2013}
{Schmieder}, B., {Guo}, Y., {Moreno-Insertis}, F., {et~al.} 2013, \aap, 559, A1

\bibitem[{{Schmieder} {et~al.}(2004){Schmieder}, {Lin}, {Heinzel}, \&
  {Schwartz}}]{Schmieder2004}
{Schmieder}, B., {Lin}, Y., {Heinzel}, P., \& {Schwartz}, P. 2004, \solphys,
  221, 297

\bibitem[{{Schmieder} {et~al.}(1988){Schmieder}, {Mein}, {Simnett}, \&
  {Tandberg-Hanssen}}]{Schmieder1988}
{Schmieder}, B., {Mein}, P., {Simnett}, G.~M., \& {Tandberg-Hanssen}, E. 1988,
  \aap, 201, 327

\bibitem[{{Schmieder} {et~al.}(1983){Schmieder}, {Mein}, {Vial}, \&
  {Tandberg-Hanssen}}]{Schmieder1983}
{Schmieder}, B., {Mein}, P., {Vial}, J.-C., \& {Tandberg-Hanssen}, E. 1983,
  \aap, 127, 337

\bibitem[{{Schmieder} {et~al.}(1995){Schmieder}, {Shibata}, {van
  Driel-Gesztelyi}, \& {Freeland}}]{Schmieder1995}
{Schmieder}, B., {Shibata}, K., {van Driel-Gesztelyi}, L., \& {Freeland}, S.
  1995, \solphys, 156, 245

\bibitem[{{Schou} {et~al.}(2012){Schou}, {Scherrer}, {Bush}, {Wachter},
  {Couvidat}, {Rabello-Soares}, {Bogart}, {Hoeksema}, {Liu}, {Duvall}, {Akin},
  {Allard}, {Miles}, {Rairden}, {Shine}, {Tarbell}, {Title}, {Wolfson},
  {Elmore}, {Norton}, \& {Tomczyk}}]{Schou2012}
{Schou}, J., {Scherrer}, P.~H., {Bush}, R.~I., {et~al.} 2012, \solphys, 275,
  229

\bibitem[{{Shibata} {et~al.}(1982){Shibata}, {Nishikawa}, {Kitai}, \&
  {Suematsu}}]{Shibata1982}
{Shibata}, K., {Nishikawa}, T., {Kitai}, R., \& {Suematsu}, Y. 1982, \solphys,
  77, 121

\bibitem[{{Shibata} {et~al.}(1992){Shibata}, {Nozawa}, \&
  {Matsumoto}}]{shibata1992}
{Shibata}, K., {Nozawa}, S., \& {Matsumoto}, R. 1992, PASJ, 44, 265

\bibitem[{{Shimojo} {et~al.}(1996){Shimojo}, {Hashimoto}, {Shibata},
  {Hirayama}, {Hudson}, \& {Acton}}]{Shimojo1996}
{Shimojo}, M., {Hashimoto}, S., {Shibata}, K., {et~al.} 1996, \pasj, 48, 123

\bibitem[{{Shimojo} \& {Shibata}(2000)}]{Shimojo2000}
{Shimojo}, M. \& {Shibata}, K. 2000, \apj, 542, 1100

\bibitem[{{Sterling} {et~al.}(2016){Sterling}, {Moore}, {Falconer}, {Panesar},
  {Akiyama}, {Yashiro}, \& {Gopalswamy}}]{Sterling2016}
{Sterling}, A.~C., {Moore}, R.~L., {Falconer}, D.~A., {et~al.} 2016, \apj, 821,
  100

\bibitem[{{Syntelis} {et~al.}(2019){Syntelis}, {Priest}, \&
  {Chitta}}]{Syntelis2019}
{Syntelis}, P., {Priest}, E.~R., \& {Chitta}, L.~P. 2019, \apj, 872, 32

\bibitem[{{T{\"o}r{\"o}k} {et~al.}(2009){T{\"o}r{\"o}k}, {Aulanier},
  {Schmieder}, {Reeves}, \& {Golub}}]{Torok2009}
{T{\"o}r{\"o}k}, T., {Aulanier}, G., {Schmieder}, B., {Reeves}, K.~K., \&
  {Golub}, L. 2009, \apj, 704, 485

\bibitem[{{Wyper} {et~al.}(2017){Wyper}, {Antiochos}, \& {DeVore}}]{Wyper2017}
{Wyper}, P.~F., {Antiochos}, S.~K., \& {DeVore}, C.~R. 2017, \nat, 544, 452

\bibitem[{{Wyper} {et~al.}(2019){Wyper}, {DeVore}, \& {Antiochos}}]{Wyper2019}
{Wyper}, P.~F., {DeVore}, C.~R., \& {Antiochos}, S.~K. 2019, \mnras, 490, 3679

\bibitem[{{Yang} {et~al.}(2020){Yang}, {Zhang}, {Xu}, {Zhang}, {Zhong}, \&
  {Guo}}]{Yang2020}
{Yang}, S., {Zhang}, Q., {Xu}, Z., {et~al.} 2020, arXiv e-prints,
  arXiv:2005.09613

\bibitem[{{Yeates} \& {Hornig}(2011)}]{Yeates2011}
{Yeates}, A.~R. \& {Hornig}, G. 2011, Journal of Physics A Mathematical
  General, 44, 265501

\bibitem[{{Yokoyama} \& {Shibata}(1996)}]{yokoyama1996}
{Yokoyama}, T. \& {Shibata}, K. 1996, Astrophysical Letters Communications, 34,
  133

\bibitem[{{Zhang} \& {Ji}(2014)}]{Zhang2014}
{Zhang}, Q.~M. \& {Ji}, H.~S. 2014, \aap, 567, A11

\bibitem[{{Zhao} {et~al.}(2017){Zhao}, {Schmieder}, {Li}, {Pariat}, {Zhu},
  {Feng}, \& {Grubecka}}]{Zhao2017}
{Zhao}, J., {Schmieder}, B., {Li}, H., {et~al.} 2017, \apj, 836, 52

\bibitem[{{Zuccarello} {et~al.}(2015){Zuccarello}, {Aulanier}, \&
  {Gilchrist}}]{Zuccarello2015}
{Zuccarello}, F.~P., {Aulanier}, G., \& {Gilchrist}, S.~A. 2015, \apj, 814, 126

\end{thebibliography}

\bibliographystyle{aa}
\end{document}